# PAColorHolo: A Perceptually-Aware Color Management Framework for Holographic Displays

CHUN CHEN, MINSEOK CHAE, and SEUNG-WOO NAM, Seoul National University, Republic of Korea
MYEONG-HO CHOI, MINSEONG KIM, and EUNBI LEE, Seoul National University, Republic of Korea
YOONCHAN JEONG* and JAE-HYEUNG PARK*, Seoul National University, Republic of Korea

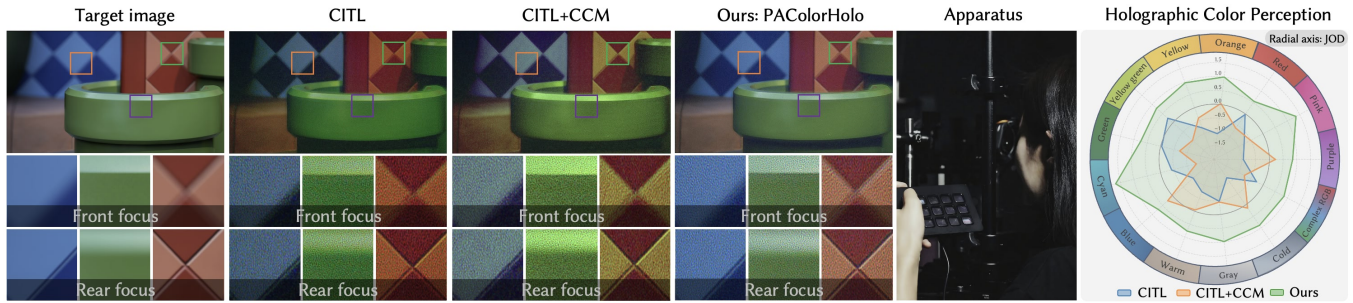

Fig. 1. Experimentally captured holographic images compared against the target reference (first column), and color perception user study results of different algorithms. Recent developments in computer-generated holography have primarily focused on reducing speckle noise and other low-level artifacts. For example, the camera-in-the-loop (CITL) method improves pixel-level fidelity by incorporating captured feedback into the hologram optimization process (second column). However, these methods often overlook perceptual color fidelity. Standard approaches for color correction, such as applying a conventional color correction matrix (CCM) calibrated with a color checker, fail to account for the narrow spectral bandwidth of laser illumination (third column). In contrast to prior methods that focus primarily on pixel-level image quality, we introduce a perceptually-aware color management framework that transforms the hologram optimization pipeline by incorporating perceptual color modeling into the rendering process (fourth column). To further validate the effectiveness of our method, we conduct a user study (fifth column). Just-Objectionable-Difference (JOD) scores are used to quantify the perceptual color performance across algorithms and are visualized by color groups using polar plots (sixth column), clearly demonstrating the perceptual color fidelity improvements achieved by our approach.

Holographic displays offer significant potential for augmented and virtual reality applications by reconstructing wavefronts that enable continuous depth cues and natural parallax without vergence–accommodation conflict. However, despite advances in pixel-level image quality, current systems struggle to achieve perceptually accurate color reproduction—an essential component of visual realism. These challenges arise from complex system-level distortions caused by coherent laser illumination, spatial light modulator imperfections, chromatic aberrations, and camera-induced color biases. In this work, we propose a perceptually-aware color management framework for holographic displays that jointly addresses input–output color inconsistencies through color space transformation, adaptive illumination control, and neural network–based perceptual modeling of the camera's color response. We validate the effectiveness of our approach through numerical simulations, optical experiments, and a controlled user study. The results demonstrate substantial improvements in perceptual color fidelity, laying the groundwork for perceptually driven holographic rendering in future systems.

*Corresponding authors

Authors' Contact Information: Chun Chen, chenchun@snu.ac.kr; Minseok Chae, mschae3d@gmail.comm; Seung-Woo Nam, 711asd@snu.ac.kr, Seoul National University, Seoul, Republic of Korea; Myeong-Ho Choi, mhchoi960905@snu.ac.kr; Minseong Kim, vkminseong@snu.ac.kr; Eunbi Lee, dldmsql0215@snu.ac.kr, Seoul National University, Seoul, Republic of Korea; Yoonchan Jeong, yoonchan@snu.ac.kr; Jae-Hyeung Park, jaehyeung@snu.ac.kr, Seoul National University, Seoul, Republic of Korea.

## 1 Introduction

Holographic displays enable realistic three-dimensional (3D) image synthesis through accurate wavefront reconstruction, providing continuous depth cues and resolving the vergence–accommodation conflict [Chang et al. 2020a; Kim et al. 2024; Maimone et al. 2017]. Beyond depth perception, such displays also support advanced functions like vision correction and optical aberration compensation [Kim et al. 2021]. These capabilities further enhance the potential of holographic displays for next-generation augmented and virtual reality (AR/VR) systems [Park and Lee 2022]. However, current hardware limitations, such as the restricted modulation capabilities of commercial spatial light modulators (SLMs) and imperfect optical systems, pose significant challenges to image quality. To address imperfections, recent research has predominantly focused on enhancing spatial fidelity through artifact reduction and noise suppression, leveraging gradient-based optimization and deep learning techniques [Choi et al. 2021; Peng et al. 2020; Shi et al. 2021].

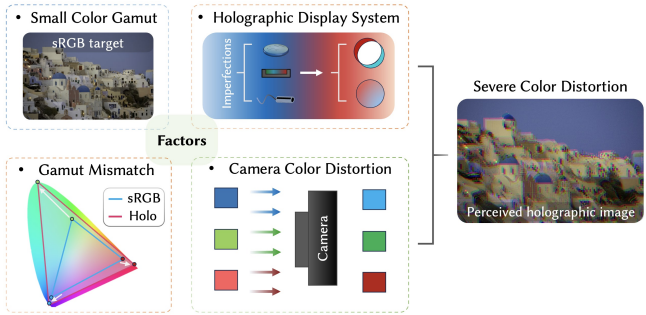

Fig. 2. Key factors affecting perceptual color fidelity in holographic display systems. During hologram optimization, mismatches between the target color space and the system's native color gamut may result in hue shifts or saturation errors. Imperfections in the optical system, such as chromatic aberrations and inaccurate illumination balancing, can introduce additional color distortions. Furthermore, the camera may contribute to spectral inaccuracies due to sensor characteristics. These intertwined factors collectively pose significant challenges to achieving perceptually accurate color reproduction. Images Credits: MIT-Adobe FiveK Dataset [Bychkovsky et al. 2011].

While these approaches have notably improved visual clarity, they typically address color fidelity only through pixel-level metrics rather than perceptual correctness. More importantly, they seldom address perceptual color fidelity from a system-level perspective, overlooking how color distortions emerge from the interplay between the system's input (e.g., laser illumination, color space matching) and output (e.g., optical system, imaging sensors). Consequently, achieving perceptually faithful color reproduction—essential for immersive holographic experiences—remains a critical challenge. This complexity is further exacerbated by the characteristics of holographic displays, which rely on coherent laser sources and the joint modulation of intensity and phase to reconstruct 3D content. This process is fundamentally different from the pixel-based, broadband emission designed for 2D panel displays in conventional systems [Huang et al. 2020; Peng et al. 2021]. As a result, standardized calibration workflows based on additive color models and precise lookup tables cannot be directly applied. Figure 2 summarizes the intertwined system-level factors that collectively pose significant barriers to accurate color reproduction. However, these intertwined challenges remain insufficiently addressed in a systematic manner.

In this study, we propose a perceptually-aware color management framework explicitly designed to enhance perceptual color fidelity in holographic displays. Unlike prior approaches that address isolated factors, our framework holistically tackles color gamut mismatches, optical system's color distortions, and camera-induced inaccuracies. Through comprehensive experiments and controlled user studies, we demonstrate substantial improvements in perceptual color fidelity. Specifically, our contributions include:

- Introducing a comprehensive perceptually-aware color management framework that significantly improves perceived color realism in holographic displays[1].

- Developing a multilayer perceptron (MLP)-based perceptual modeling approach for camera color restoration tailored to holographic display conditions.
- Designing and conducting controlled user studies to quantitatively validate improvements in perceptual color fidelity for holographic displays.
- Providing a practical pipeline for generating perceptually corrected holographic datasets, serving as potential training data for future real-time, learning-based holographic image generation methods.

## 2 Related Work

### 2.1 Computer-Generated Holography

Computer-Generated Holography (CGH) reconstructs realistic 3D scenes by precisely modulating light wavefronts. This control facilitates depth manipulation, aberration compensation, and vision correction, paving the way for next-generation near-eye display architectures [Chen et al. 2021; Jang et al. 2018; Kim et al. 2021; Maimone et al. 2017]. For instance, CGH facilitates compact AR in waveguide-based displays via accurate propagation modeling [Gopakumar et al. 2024; Jang et al. 2024; Kim and Park 2018; Peng et al. 2020], while also enabling Maxwellian-view systems to extend depth of field through computational wavefront control [Chang et al. 2019; Park and Kim 2018]. These implementations demonstrate the capacity of CGH to shift hardware-bound optical functions into the computational domain [Chae et al. 2025]. To optimize the hologram generation process, research has evolved from traditional iterative algorithms [Fienup 1982; Wu et al. 2021b]—which often suffer from slow convergence—to advanced gradient-based approaches [Chen et al. 2024a; Nam et al. 2023; Peng et al. 2020] and learned holography techniques [Shi et al. 2021; Wu et al. 2021a]. These modern methods leverage differentiable loss functions and deep neural networks to achieve real-time, high-fidelity synthesis [Gopakumar et al. 2024; Peng et al. 2020]. However, despite these advancements, existing approaches predominantly focus on numerical fidelity and wave propagation, often overlooking the critical role of human visual perception in holographic display quality.

### 2.2 Human-Centered Holography

Recent human-centered holography shifts optimization toward the human visual system (HVS). Importantly, the perceptual quality of a hologram depends not only on the fidelity of reconstructed wavefronts but also on how the eye responds to depth cues, parallax, and viewing comfort constraints [Chang et al. 2020a]. To address these concerns, several HVS-inspired strategies have emerged. Pupil-considered holography adapts generation to eye dynamics, aligning wavefronts with the entrance pupil to improve angular resolution and brightness [Chakravarthula et al. 2022a; Wang et al. 2024b]. Similarly, foveated holography [Cem et al. 2020; Chang et al. 2020b; Ju and Park 2019; Kim et al. 2019] exploits the decline in peripheral acuity, allocating computational resources where perception is most sensitive [Chakravarthula et al. 2021; Patney et al. 2016]. To further enhance the viewing experience, field-of-view (FOV) and eye-box expansion techniques [Chae et al. 2023; Jang et al. 2018; Li et al. 2022; Tseng et al. 2024; Zheng et al. 2025] extend visual coverage and accommodate natural head and eye movements, thereby reducing alignment sensitivity and visual fatigue [Chang et al. 2019; Choi et al. 2020; Kim and Park 2018]. Furthermore, recent efforts have enhanced accommodation response [Kim et al. 2022; Lee et al. 2024; Yang et al. 2022] and employed time-multiplexing [Choi et al. 2022] or light-field approaches [Chakravarthula et al. 2022b; Kim et al. 2024; Park and Askari 2019]. These methods reproduce natural depth cues consistent with physiological conditions [Chang et al. 2020a; Kramida 2015], effectively mitigating eye strain. As display specifications advance, integrating human factors into CGH becomes not merely an enhancement but a necessity, rendering perceptual optimization as critical as hardware improvements.

---
[1]Project page: https://3dholo.github.io/PAColorHolo/

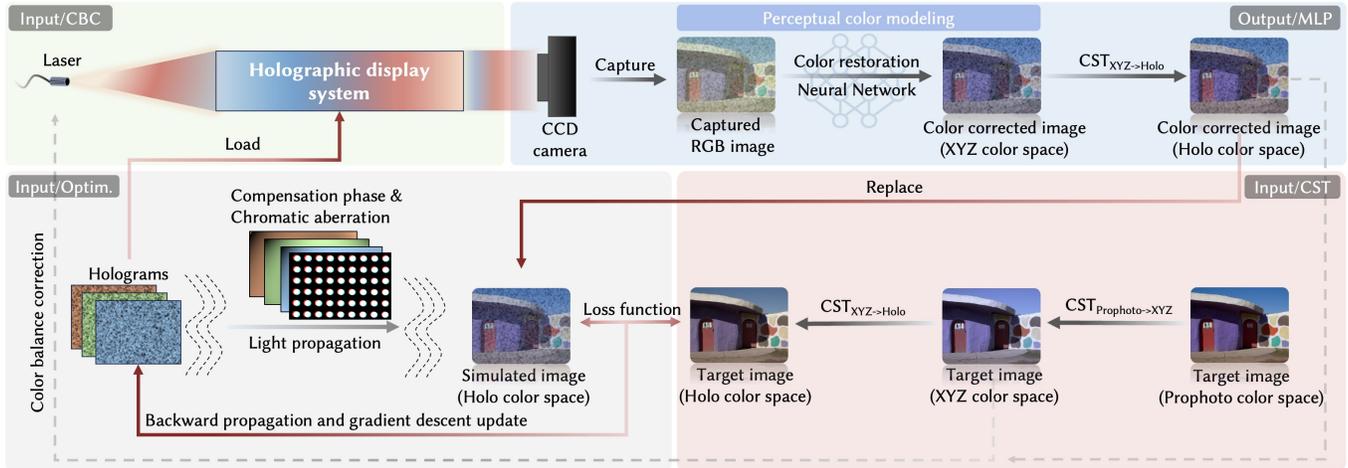

Fig. 3. Schematic of the proposed perceptually-aware color management framework. The holographic display system is organized into input and output modules to facilitate precise color control. The input module consists of three key components: color balance correction (CBC) for laser illumination, hologram optimization, and color space transformation (CST) of the target image. The output module employs a trained MLP to restore perceptual color fidelity by mapping the camera-captured RGB values to the CIE XYZ color space. The RGB power ratio of the laser is first initialized based on the transformed target image and is subsequently refined using feedback from the captured image. During light propagation simulation, phase compensation and inverse homography correction are applied to mitigate optical and chromatic aberrations. The MLP output, represented in the CIE XYZ color space, is subsequently converted into the Holo color space. This allows for camera-integrated optimization, where the simulated reconstruction is replaced with the color-corrected image captured by the camera. In the optimization loop, the target image is first transformed into the Holo color space to ensure consistency with the propagation output. A loss is then computed between the color-corrected reconstruction and the target image, and gradients are derived via backpropagation to iteratively update the holograms using gradient descent. Through this pipeline, the proposed framework effectively enhances perceptual color fidelity in the reconstructed holographic images. Images Credits: MIT-Adobe FiveK Dataset [Bychkovsky et al. 2011].

### 2.3 Perceptual Color Challenges

While HVS-driven optimizations for spatial fidelity and depth cues have been extensively explored, color perception remains a critical yet underexplored determinant of visual realism and immersion [Gil Rodríguez et al. 2022; Witt 1995; Zhang et al. 2021]. Unlike conventional displays that rely on standardized calibration for incoherent light [Bastani et al. 2005; Consortium et al. 2010; Reinhard and Urban 2022; Sharma 2002], holographic systems face unique distortions arising from coherent laser illumination, complex wavefront encoding and optical system imperfections [Peng et al. 2020]. These fundamental differences necessitate color management strategies specifically tailored to the physics of holographic displays.

To mitigate these optical distortions and hardware constraints, research has explored aberration correction [Kim et al. 2021; Nam et al. 2020], SLM compensation [Li and Cao 2019; Shi et al. 2022], and high-dynamic-range (HDR) techniques [Kadis et al. 2022; Kavaklı et al. 2023; Yonesaka et al. 2016]. Additionally, wavelength multiplexing has been employed to expand the achievable color gamut and improve image clarity [Schiffers et al. 2025; Trisnadi 2002]. Recently, camera-integrated methods have emerged to further mitigate optical system imperfections and improve holographic image quality via real-world feedback [Chakravarthula et al. 2020; Chao et al. 2023; Chen et al. 2022; Wang et al. 2024a; Xia et al. 2025]. However, these methods heavily rely on the accuracy of the captured images, for which precise color representation is not guaranteed. Traditional calibration techniques, such as those using color checkers and color correction matrices (CCMs) [Finlayson et al. 2015], often perform poorly under narrowband and coherent laser illumination. This limitation arises from spectral mismatches between the laser primaries and the reflectance properties of the calibration targets [Li et al. 2023; Molada-Tebar et al. 2024]. Measuring the camera's spectral sensitivity curve has been explored as a potential solution [Li et al. 2023]; however, this approach typically requires complex, optically sealed setups and often overlooks the influence of internal optoelectronic processing [Gong et al. 2022; Jia et al. 2024; Martínez-Verdú et al. 2002; Solomatov and Akkaynak 2023].

Consequently, current methods predominantly focus on pixel-level fidelity rather than perceptual color accuracy. To bridge this gap, we introduce a perceptually-aware optimization framework to mitigate the perceptual color discrepancy between rendered holographic images and human visual expectations. We further validate our approach through controlled user studies that quantitatively assess perceptual color fidelity. We expect our framework to offer a practical and effective solution for perceptually accurate color reproduction in real-world holographic display systems.

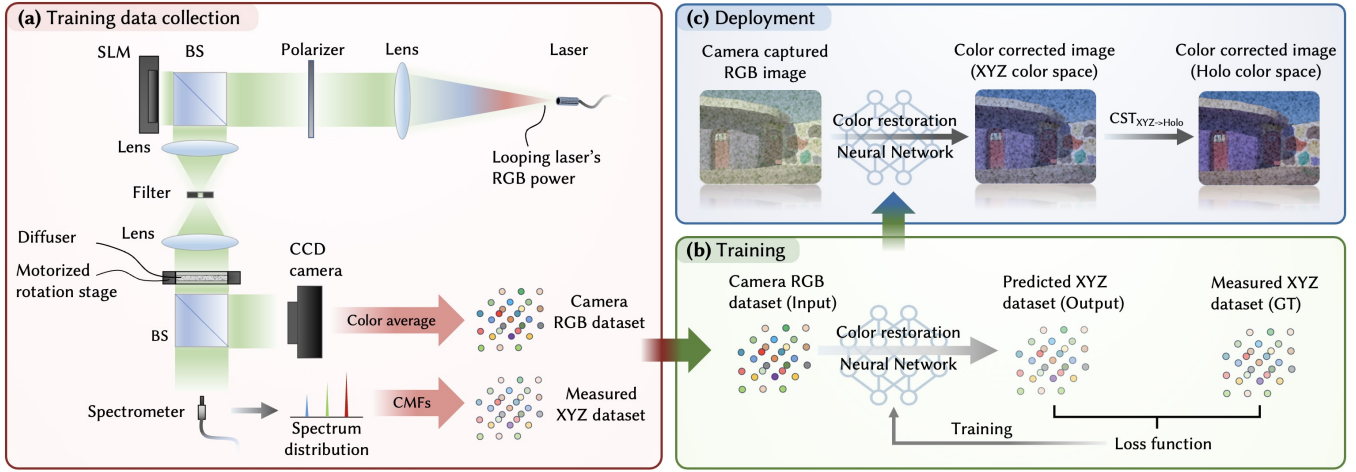

Fig. 4. Overview of the color restoration MLP framework. (a) Data acquisition pipeline for capturing camera RGB responses and corresponding spectral measurements under a holographic display setup. (b) The resulting dataset, consisting of paired camera RGB and measured XYZ values, is used to train an MLP that maps RGB inputs to ground-truth XYZ outputs. (c) The trained MLP restores the accurate color representation of captured images in the XYZ color space, which is then converted to the Holo color space using a $CST\{\cdot\}$ operation. Images Credits: MIT-Adobe FiveK Dataset [Bychkovsky et al. 2011].

## 3 Perceptually-Aware Hologram Optimization and Color Management Framework

We present a modular three-stage color management framework designed to improve the perceptual color fidelity of holographic images. We strategically decompose the pipeline, rather than formulating it as a unified differentiable mapping, to decouple standard colorimetric conversions from device-specific corrections. This decomposition ensures that standard transformations remain mathematically precise rather than being approximated by a neural network. As illustrated in Fig. 3, our approach jointly addresses color handling at both the input and output stages. The framework consists of: (i) color space transformation (CST) to align both reconstructed and target images in a common color space; (ii) perceptual color restoration using an MLP model to correct camera-induced distortions; and (iii) color balance correction (CBC) to fine-tune the final output for perceptual consistency. By integrating these components into the hologram optimization pipeline, the proposed framework systematically addresses perceptual color fidelity challenges across the stages of hologram generation, propagation, and display. Ultimately, this modularity allows each stage to tackle a clearly defined sub-problem, paving the way for easier integration with standard graphics rendering pipelines.

### 3.1 Holographic Color Space Transformation

Images in standard RGB color spaces such as sRGB or AdobeRGB cannot be directly used as targets for holographic displays. This is because these standard spaces typically assume primaries with broader spectral bandwidths, which fundamentally differ from the narrowband, highly saturated spectral characteristics of laser-based holographic systems. This spectral mismatch in color gamuts often leads to color distortions such as hue shifts and either under- or over-saturation in the reconstructed holographic images.

To address this issue, we define a holographic color space (Holo color space) as the native device-dependent RGB space of our system. This color space explicitly reflects the spectral power distributions (SPDs) of the laser primaries used in our system. This space is constructed by integrating the measured SPDs with standard CIE 1931 color matching functions (CMFs) [Fairman et al. 1997], from which the tristimulus values of the primaries are derived. We adopt the widely used D65 illuminant as the white reference, yielding a color space $C_{Holo}$ characterized by its device-specific primaries and standard white point. This definition establishes the precise target domain for our subsequent CST stage.

The transformation between linear Holo color space and CIE XYZ color space is represented by a calibrated 3×3 matrix $\mathbf{M}_{\text{RGB}\to\text{XYZ}}$ and its inverse $\mathbf{M}_{\text{XYZ}\to\text{RGB}}$. These matrices are derived from the primary tristimulus values and scaled to the D65 white point. Additional implementation details are provided in the supplementary material.

To formalize CST, we define a $CST\{\cdot\}$ operator that maps an image from a source color space $C_s$ to a target color space $C_t$:

$$I_t = CST\{I_s; C_s, C_t\} = \mathbf{M}_{\text{XYZ}\to C_t}\, \mathbf{A}_{w_s \to w_t}\, \mathbf{M}_{C_s \to \text{XYZ}}\, I_s, \quad (1)$$

where $I_s$ and $I_t$ denote the linearized RGB values of the image in the source and target color spaces, and $\mathbf{A}_{w_s \to w_t}$ is an optional chromatic adaptation matrix, which becomes the identity matrix when the white points of the two color spaces are the same. Similarly, when either $C_s$ or $C_t$ corresponds to the CIE XYZ color space, the associated transformation matrix $\mathbf{M}$ also becomes an identity matrix. In addition, we apply gamut mapping instead of hard clipping after the linear transformation when converting from a larger color gamut to a smaller color gamut to preserve color details, especially near gamut boundaries [Le et al. 2023]. Through this process, the target image can be transformed from another color space (e.g., XYZ) into the Holo color space: $I_{Holo} = CST\{I_{XYZ}; C_{XYZ}, C_{Holo}\}$.

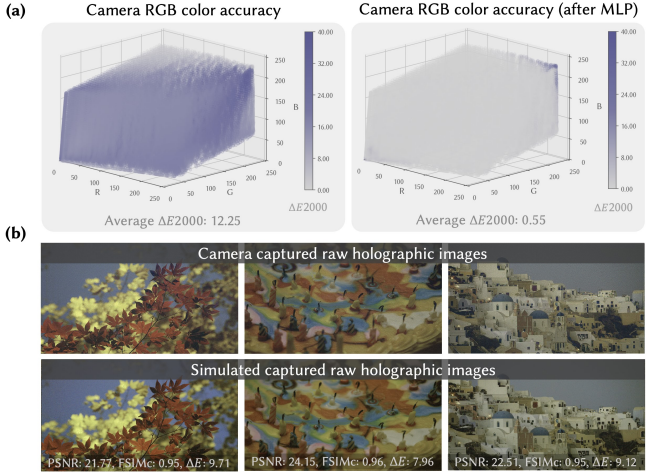

Fig. 5. Evaluation of color restoration accuracy of our trained MLP. (a) Color accuracy comparison visualized using $\Delta E_{2000}$ before and after MLP color restoration. A significant reduction in $\Delta E_{2000}$ is observed after applying the MLP, indicating improved perceptual color fidelity. Note that the remaining elevated errors in the high-intensity regions (top-right) are attributed to information loss from sensor saturation. (b) Validation of the inverse MLP through comparison between real and simulated camera captures, confirming its ability to model the camera's color response. Images Credits: MIT-Adobe FiveK Dataset [Bychkovsky et al. 2011].

### 3.2 Light Propagation Model

Optical holographic display systems inevitably introduce optical imperfections during wavefront propagation, including structural noise, optical aberrations, and chromatic aberrations. To mitigate their impact on the color fidelity of the reconstructed holographic image, we augment the wave propagation model with an additional compensation phase term $\phi_{c_i}$ and a geometric transformation operator $PT\{\cdot\}$.

The compensation phase $\phi_{c_i}$ is generated from system-specific optical aberrations, which are measured and parameterized using Zernike polynomials [Kim et al. 2021]. The perspective transformation $PT\{\cdot\}$ simulates the system's chromatic aberration by applying an inverse homography calibrated across color channels. Further implementation details and calibration procedures are provided in the supplementary material.

We employ the angular spectrum method to model light propagation in our holographic display system. After incorporating optical imperfections, the light propagation operator is defined as:

$$\mathcal{P}(\phi_i) = PT\left\{\mathcal{F}^{-1}\left\{\mathcal{F}\left\{e^{j(\phi_i+\phi_{c_i})}\right\}\cdot H(f_x, f_y, \lambda_i)\right\}\right\},$$
$$H(f_x, f_y, \lambda_i) = \begin{cases} e^{jk_iz\sqrt{1-(\lambda_if_x)^2-(\lambda_if_y)^2}}, & \text{if } \sqrt{f_x^2+f_y^2} < \frac{1}{\lambda_i},\\ 0, & \text{otherwise},\end{cases} \quad (2)$$

where $\phi_i$ is the input phase profile at wavelength $\lambda_i$, and $\mathcal{P}(\phi_i)$ denotes the resulting complex field at the propagation plane. $\phi_{c_i}$ is a compensation phase that accounts for measured system aberrations. $\mathcal{F}\{\cdot\}$ and $\mathcal{F}^{-1}\{\cdot\}$ represent the forward and inverse Fourier transforms, respectively. $H(f_x, f_y, \lambda_i)$ is the transfer function under the angular spectrum approximation, with spatial frequencies $f_x, f_y$, and $k_i = 2\pi/\lambda_i$ being the wavenumber. The propagation distance is denoted by $z$.

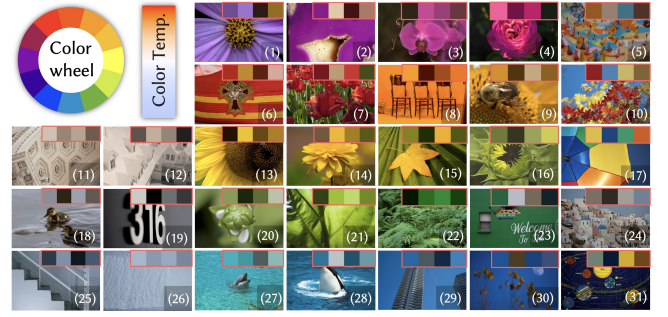

Fig. 6. Image dataset used in our experiments. The collected images are categorized by hue and color temperature. Dominant colors are extracted using color clustering to enhance visual representation. Images Credits: MIT-Adobe FiveK Dataset [Bychkovsky et al. 2011].

### 3.3 Perceptual Color Modeling Using an MLP Network

Conventional camera color calibration using a color checker and a CCM assumes broadband illumination and linear color response [Jia et al. 2024]. These assumptions often fail under narrowband lighting, making it difficult to correct the camera's inherent non-linear color distortions and leading to perceptual mismatches. Even with polychromatic illumination strategies [Schiffers et al. 2025] that introduce spectral diversity, the resulting spectrum typically remains composed of discrete peaks rather than a continuous broadband spectrum. Consequently, the sensor's integration of these discrete signals retains complex non-linearities that simple linear models cannot fully resolve. To address this issue, we propose a perceptual color restoration framework based on a fully connected MLP, leveraging its capacity to model non-linear color mappings inherent to discrete-spectrum illumination, whether in standard narrowband or polychromatic configurations. The network, denoted as $MLP\{\cdot\}$, maps camera-captured RGB values to perceptually accurate representations in the CIE XYZ color space. Figure 4 illustrates the complete workflow of data acquisition, network training, and deployment in scenarios involving narrowband illumination.

*Color Data Collection.* Our holographic display system utilizes narrowband RGB laser illumination, necessitating precise spectral and colorimetric characterization. To accurately capture the camera's color response, we collected comprehensive datasets encompassing both camera RGB values and corresponding spectral measurements, as depicted in Fig. 4(a). A rotating diffuser was placed before the camera and spectrometer to mitigate speckle noise. Furthermore, time-multiplexing was applied to stabilize the acquisition of RGB and spectral data, further enhancing measurement quality. A total of $32(R) \times 36(G) \times 39(B) = 44,928$ data pairs were acquired to construct a relatively dense dataset that adequately covers the color space. The camera-captured RGB data were averaged and normalized to a $[0, 1]$ range, while spectral data were converted to XYZ values using the CIE 1931 CMFs and normalized accordingly.

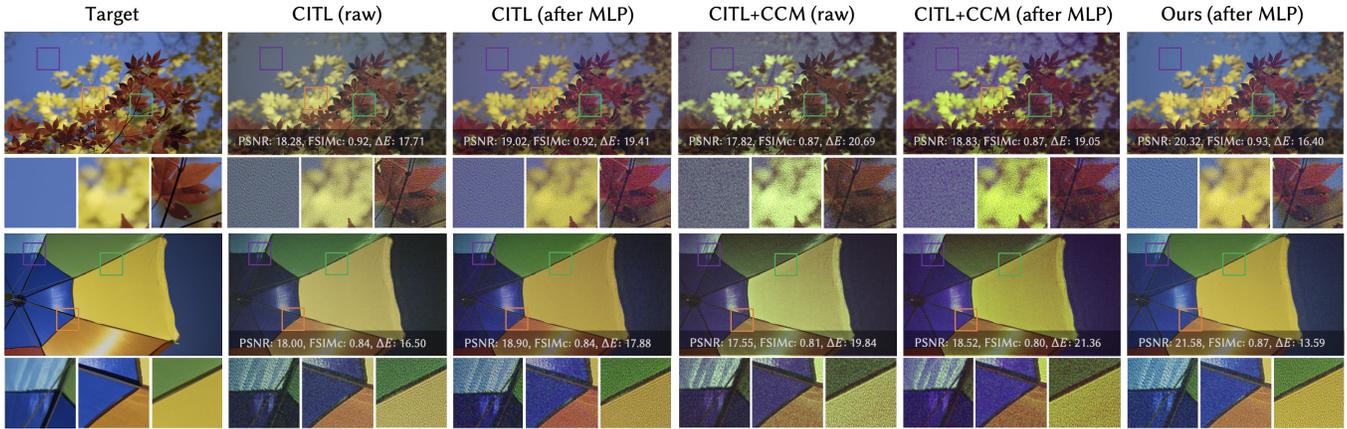

Fig. 7. Experimentally captured 2D holographic images of complex color scenes. Raw and MLP-restored images produced by different optimization methods are shown. The captured raw images from the CITL method exhibit low speckle noise (second column). However, due to the lack of a comprehensive color management process, the restored images (third column) show clear color deviations from the target (first column). The CITL+CCM method (fourth and fifth columns) integrates a CCM-based color correction into the optimization pipeline, but its performance is limited under narrowband laser illumination, leading to suboptimal color fidelity. Our proposed perceptually-aware color management framework achieves the best color reproduction performance (sixth column) compared to other methods. Images Credits: MIT-Adobe FiveK Dataset [Bychkovsky et al. 2011].

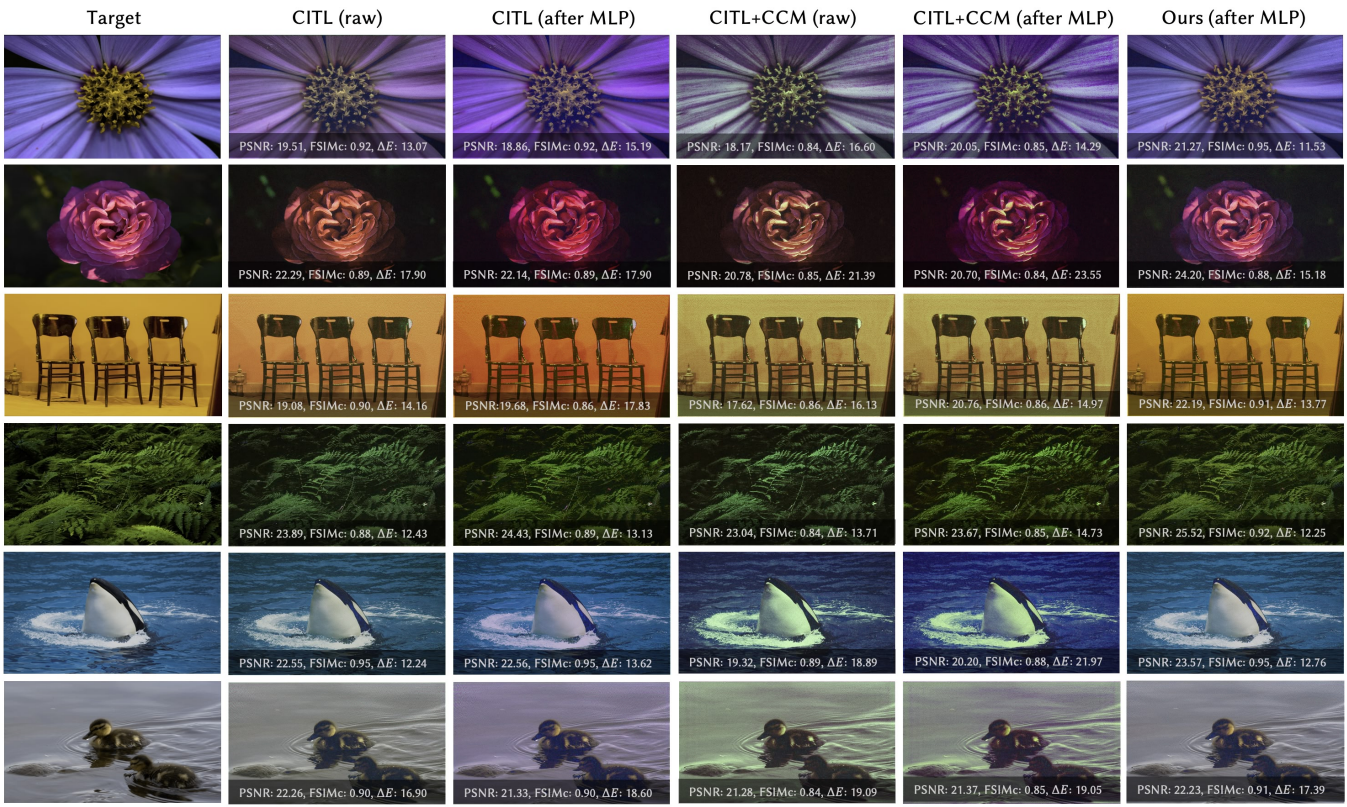

Fig. 8. Experimentally captured 2D holographic images for simple color scenes. Representative examples from the 31-scene captured holographic images are shown; full results are provided in the supplementary material. These color wheel–like holographic images facilitate the evaluation of color fidelity and image quality across different methods and color groups. Consistent with the results from complex RGB scenes, our method demonstrates superior color reproduction performance. Images Credits: MIT-Adobe FiveK Dataset [Bychkovsky et al. 2011].

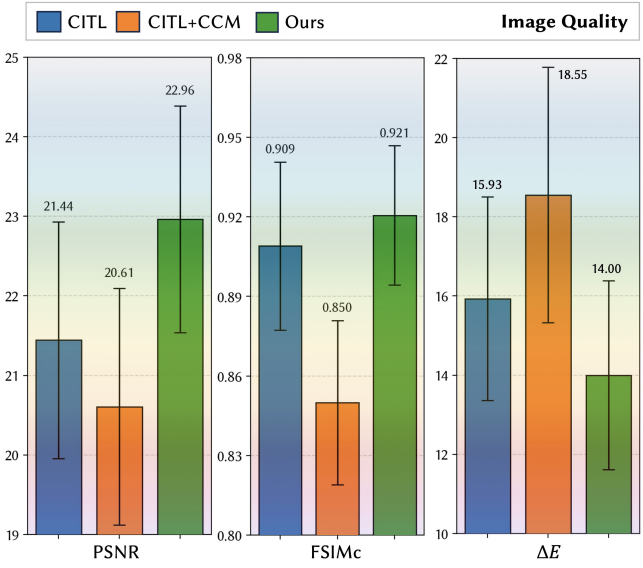

Fig. 9. Averaged image quality metrics of experimentally captured holographic images across 31 scenes, evaluated using PSNR, FSIMc (higher is better) and $\Delta E$ (lower is better).

*Color Restoration MLP Training.* The MLP network architecture consists of an input layer (dimension 3), followed by hidden layers of sizes 256, 128, 64, and 32, each with ReLU activations, and a 3-dimensional output layer. This design effectively models the nonlinear color mappings induced by the camera's internal processing in the holographic display scenario. The training pipeline is illustrated in Fig. 4(b).

Prior to training, saturated pixels (channel values exceeding a threshold of 235) and black samples (all illumination levels at zero) were excluded. The remaining data were split into monochromatic and polychromatic subsets, both used concurrently to ensure robustness across varying illumination conditions. We trained the network via supervised regression, minimizing the L1 distance between predicted and ground-truth XYZ values. We utilized the AdamW optimizer [Loshchilov and Hutter 2017] with a learning rate of 0.0015, weight decay of $10^{-5}$, and cosine annealing over 2000 epochs, with a minimum learning rate of zero. During each training epoch, mixed batches from both the monochromatic and polychromatic subsets were processed. Validation was performed using both the L1 loss and the perceptual color difference metric $\Delta E$ in CIE Lab space [Zhang et al. 2021], where a lower $\Delta E$ indicates higher color accuracy. The final model was selected based on its overall validation performance. Once trained, the MLP can be applied to camera-captured images to restore their color appearance in the CIE XYZ space. This restored representation can then be converted to other color spaces via the $CST\{\cdot\}$ operation, as illustrated in Fig. 4(c).

*Evaluation of MLP Color Accuracy.* We evaluated the MLP's color accuracy by converting camera-captured RGB values into the Holo color space and computing their color difference $\Delta E$. As shown in Fig. 5(a), the left column visualizes the raw color discrepancies, with

**Algorithm 1** PAColorHolo Optimization Process

1: $R(u, v)$ : replace values of $u$ with $v$ while retaining gradients from $u$ for backpropagation
2: $I_c, I_o$ : captured and target images in their original color spaces
3: $I_{tholo}, I_{cholo}$ : target and captured images in the Holo color space
4: $\mathcal{P}(\phi)$ : light propagation model with input phase $\phi$
5: $s$ : learnable energy scaling factor
6: $\alpha_\phi, \alpha_s$ : learning rates for phase and scaling factor updates
7: $K$ : number of optimization iterations
8: $\mathcal{L}$ : mean squared error (MSE) loss function
9: **Input:** $I_{tholo} = CST\{I_o; C_o, C_{holo}\}$   ▷ to Holo color space
10: **Initialize:** Phase profile $\phi \leftarrow \phi_0$    ▷ random initialization
    Scaling factor $s \leftarrow s_0 = 1$   ▷ constant initialization
    $r, g, b \leftarrow CBC\{CST\{I_{tholo}; C_{holo}, C_{XYZ}\}\}$ ▷ laser ini.
11: **for** $k$ **in** $1 \ldots K$ **do**
12:   $I_{cholo} = CST\{MLP\{I_c(\phi)\}; C_{XYZ}, C_{holo}\}$  ▷ to Holo color space
13:   $\phi \leftarrow \phi - \alpha_\phi \nabla_\phi \left( \mathcal{L}\left( s \cdot R(|\mathcal{P}(\phi)|^2, I_{cholo}), I_{tholo} \right) \right)$
14:   $s \leftarrow s - \alpha_s \nabla_s \left( \mathcal{L}\left( s \cdot R(|\mathcal{P}(\phi)|^2, I_{cholo}), I_{tholo} \right) \right)$
15:   **if** $k = K/2$ **then**
16:     $r, g, b \leftarrow CBC\{CST\{I_{cholo}; C_{holo}, C_{XYZ}\}\}$ ▷ laser update
17:   **end if**
18: **end for**
19: **return** $\phi, s$

deeper hues indicating lower accuracy. The right column demonstrates the corrected output, where the trained MLP significantly reduces color errors.

While these results confirm the MLP's ability to approximate spectrally measured colors, they do not fully validate its effectiveness in restoring the camera-perceived colors. To address this, we introduced an inverse MLP, mirroring the architecture of the original MLP but trained with reversed input-output pairs. This inverse model simulates the camera's color response by mapping spectral data back to RGB values. In our simulation, holographic images (converted to XYZ space) served as inputs to the inverse MLP, and the output approximated the camera's raw RGB responses. Figure 5(b) compares real and simulated raw captures, demonstrating that the inverse MLP effectively models the camera's color response and further supporting the MLP's effectiveness in restoring camera-perceived color.

### 3.4 Two-Step Color Balance Correction

In addition to correcting perceptual color distortion on the camera side, our framework also manages the illumination side, specifically the RGB power ratio of the laser, which directly influences the hologram optimization convergence and the perceived color distribution in the final holographic image. Here, we introduce a two-step, perceptually guided CBC strategy that explicitly regulates the energy distribution of the laser to better align with the target image's color composition.

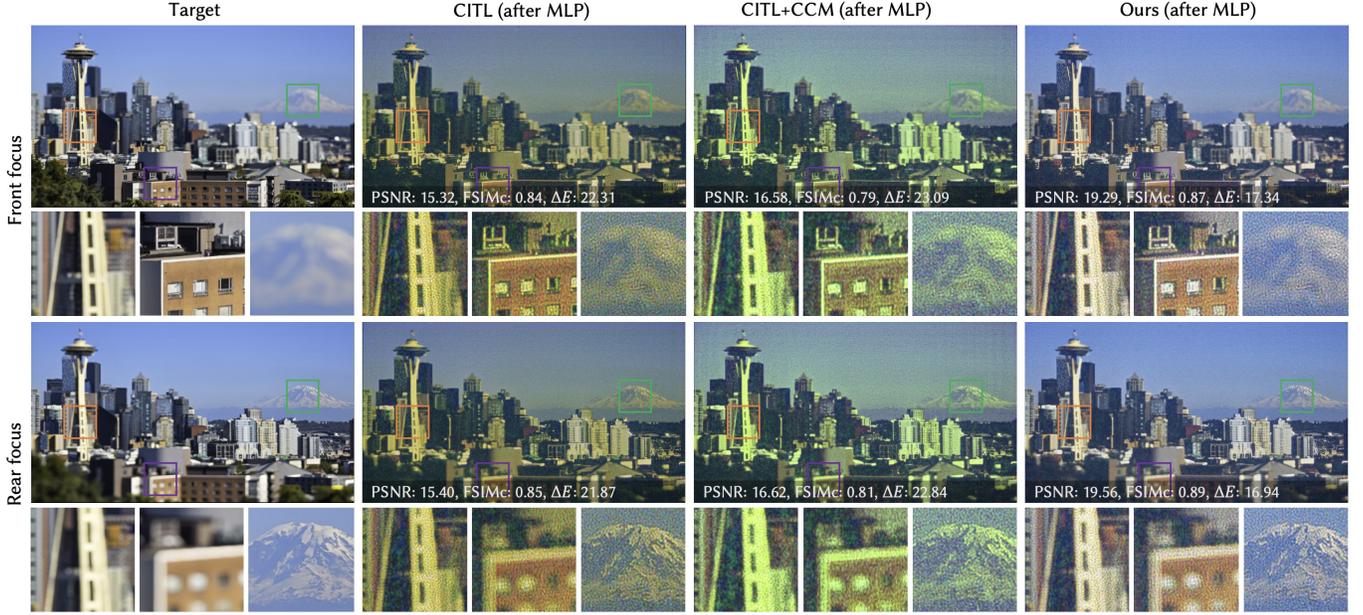

Fig. 10. Experimentally captured 3D holographic images of complex color scenes. Using the 3D focal stack as the target (first column), all methods exhibit natural defocus blur. However, without proper color management, the CITL (second column) and CITL+CCM (third column) methods display color shifts towards warm and green hues, respectively, deviating from the target. Our proposed color management framework achieves the best color fidelity and consistency across different depths (fourth column).

In the first stage, the initial laser RGB power ratio is estimated directly from the target image's average XYZ color using the CMFs. This laser RGB power ratio computation is denoted as $CBC\{\cdot\}$. In the second stage, the laser RGB power ratio is refined based on the color information of the camera-captured image, which is first transformed to the XYZ color space via the $MLP\{\cdot\}$ operation. The resulting average XYZ value is then used to adjust the laser RGB power ratio with the $CBC\{\cdot\}$ operation. This secondary adjustment is strategically applied at the midpoint of the optimization, where the energy distribution stabilizes. While additional adjustment steps are possible, they may compromise the stability of the optimization. The proposed two-step CBC approach significantly improves the color fidelity of the reconstructed holographic images. Further details of this process are provided in the supplementary material.

### 3.5 Optimization Process

Our hologram optimization framework conceptualizes the display system as comprising two coordinated components: input-side and output-side color management modules. The input module includes the $CBC\{\cdot\}$ operation, the propagation model $\mathcal{P}\{\cdot\}$, and the color space transform $CST\{\cdot\}$. The output module applies the $MLP\{\cdot\}$ to restore perceptually accurate colors from camera-captured images. The overall optimization aims to determine the optimal full-color phase profile $\phi$ and a global energy scaling factor $s$ that minimize the discrepancy between the reconstructed holographic image and the target image, which is formulated as the following minimization problem:

$$\arg\min_{\phi,s} \mathcal{L}\left(s \cdot R(|\mathcal{P}(\phi)|^2, I_{cholo}), I_{tholo}\right). \qquad (3)$$

The complete optimization process is summarized in Algorithm 1. In this optimization framework, the mean squared error (MSE) loss function $\mathcal{L}$ quantifies the difference between the reconstructed holographic image $I_{cholo}$ and the target image $I_{tholo}$, computed in the linear Holo color space. Here, $R$ denotes a replacement operation that substitutes the simulated result with the camera-captured image $I_{cholo}$ during the forward pass, while propagating gradients through the simulation model during backpropagation. The optimization leverages the color-field-sequential (CFS) operation of the SLM, synchronized with the RGB laser system, ensuring full-color holographic reconstruction with simultaneous RGB hologram optimization.

To accelerate convergence and enhance stability, a learnable energy scaling factor $s$ is introduced, which modulates the intensity of the captured full-color holographic image. The optimization is carried out using the Adam optimizer [Kingma and Ba 2014], which jointly updates both the phase profile $\phi$ and the scaling factor $s$ over $K$ iterations.

## 4 Holographic Reconstruction Verification

### 4.1 Experimental System Setup

We construct a full-color holographic display system based on an RGB fiber-coupled laser module and a phase-only SLM. The red, green, and blue lasers operate at 636 nm, 512 nm, and 453 nm, respectively. The SLM provides 1920 × 1080 resolution with 6.4 μm pixel pitch and is driven in CFS mode at 60 Hz. A 4-$f$ optical system is employed for noise filtering, and the holographic image is reconstructed at a propagation distance of 120 mm. A CCD camera (without a lens), synchronized with the display system, captures the raw full-color holographic images (only demosaiced). Further implementation details are provided in the supplementary material.

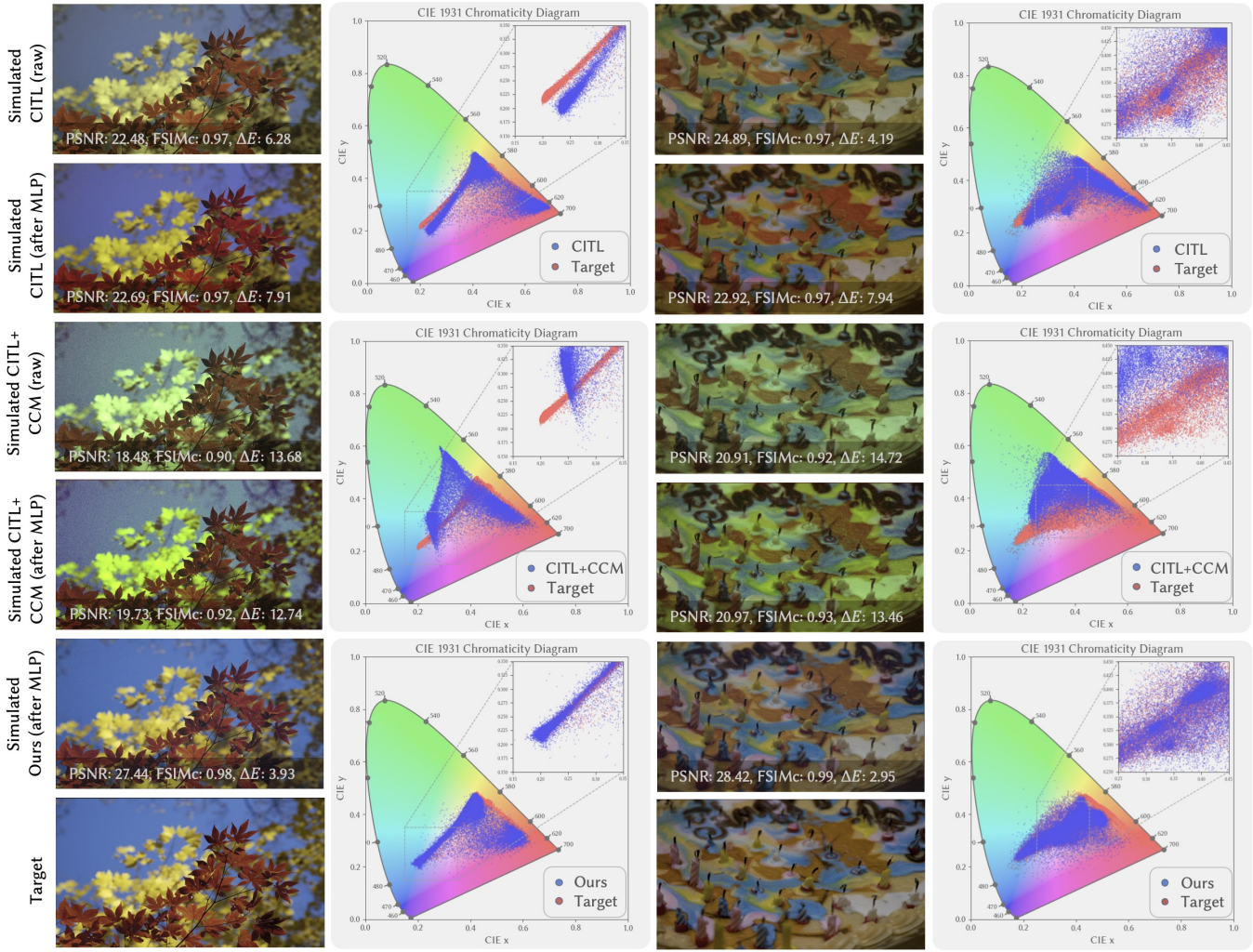

Fig. 11. Simulated holographic images and corresponding chromaticity diagram comparisons across different algorithms. Red markers indicate the target image's colors, while blue markers represent the colors reproduced by each algorithm. For the CITL method, blue colors shift towards purple-blue, and brown hues shift towards red-orange, as evidenced by the reconstructed images and chromaticity plots (first and second rows). The CITL+CCM method exhibits further degraded color fidelity, introducing a prominent green cast (third and fourth rows). In contrast, our proposed method achieves the best color reproduction, with most colors overlapping with the target and superior image quality (fifth row). Images Credits: MIT-Adobe FiveK Dataset [Bychkovsky et al. 2011].

In simulation experiments, we replicate the SLM parameters, RGB laser wavelengths, and propagation distance to mirror the optical setup. To account for illumination imperfections, we introduce controlled intensity variations in the light source. The two-step color balance correction is also applied using the average colors of the target and reconstructed images. The trained inverse MLP simulates the camera-captured raw holographic image, while the MLP restores the simulated raw image to the XYZ color space. The remaining optimization procedures follow the same protocol as the optical experiment, aiming to closely approximate the algorithm's behavior in the physical system.

### 4.2 Dataset Design

For both optical and user study experiments, we construct a curated subset of 31 ProPhoto RGB images selected from the MIT-Adobe FiveK Dataset [Bychkovsky et al. 2011]. The images are categorized into groups based on hue and color temperature. For each image, dominant colors are extracted using the Mini-Batch K-Means algorithm [Peng et al. 2018; Xiao et al. 2018], as shown in the highlighted regions of Fig. 6. This classification facilitates a comprehensive evaluation of the perceptual color fidelity across different algorithms and various color categories.

Table 1. Ablation study of different component combinations on simulated holographic images, reporting average $\Delta E$, PSNR, and FSIMc across 31 test scenes.

| Method | $\Delta E\downarrow$ | PSNR (dB)$\uparrow$ | FSIMc$\uparrow$ |
| --- | --- | --- | --- |
| CITL | 9.11 | 22.50 | 0.96 |
| + CST | 9.14 | 22.29 | 0.96 |
| + CBC | 7.81 | 23.61 | 0.95 |
| + MLP | 8.06 | 23.97 | 0.96 |
| + MLP + CBC | 6.07 | 25.98 | 0.96 |
| + CST + MLP | 7.37 | 24.48 | 0.96 |
| + CST + CBC | 8.05 | 23.40 | 0.95 |
| + CST + MLP + CBC (Ours) | **5.39** | **27.08** | **0.97** |

### 4.3 Experimental Results

We compare our method with the state-of-the-art hologram optimization technique, the camera-in-the-loop (CITL) method. We also evaluate an enhanced variant, CITL+CCM, which applies a color checker–optimized CCM to correct captured images during optimization. The pseudocode for all compared algorithms is provided in the supplementary material. Qualitative comparisons based on camera-captured images are presented in Figs. 7 and 8. Note that "after MLP" denotes that the captured raw image is processed by the color restoration MLP and transformed into the Holo color space via a CST operation. This step aims to approximate the perceptual target colors more accurately, thereby providing a clearer basis for comparing the color reproduction performance across algorithms.

Since all methods adopt the same propagation model, system-specific optical and chromatic aberrations are consistently mitigated. Moreover, noise artifacts are effectively suppressed across all approaches via camera-integrated optimization. However, when evaluating color fidelity, both the raw outputs and the MLP-restored images from the CITL and CITL+CCM methods exhibit noticeable color deviations when compared to the target images. This degradation stems from the fact that neither method effectively addresses the complex, non-linear color distortions introduced by the camera and optical system under narrowband illumination conditions. In contrast, our method explicitly incorporates perceptual camera color restoration, color balance correction, and color space alignment, which jointly contribute to the significantly improved color fidelity. Figure 9 summarizes the average image quality metrics across 31 scenes, evaluated using peak signal-to-noise ratio (PSNR), $\Delta E$, and FSIMc [Zhang et al. 2011]. The results consistently demonstrate that our approach outperforms other methods.

To further validate our method, we captured 3D focal stack images for complex color scenes, as illustrated in Fig. 10. Consistent with the trends observed in 2D evaluations, the algorithms exhibit similar color reproduction characteristics in the 3D context. Specifically, the CITL method tends to produce a warmer color tone, while the CITL+CCM tends towards a greenish tint. Our method maintains superior color fidelity across depths, demonstrating robust performance in complex 3D scenes.

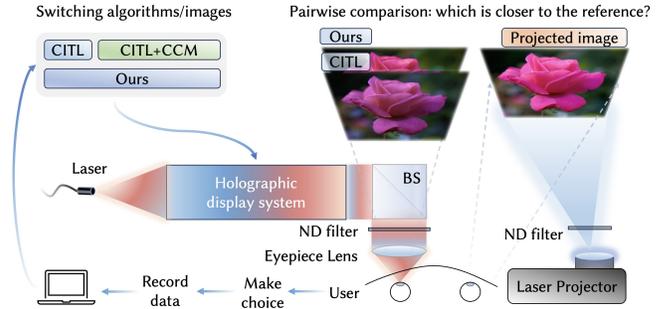

Fig. 12. Schematic of the designed user study system for the holographic color perception evaluation. Images Credits: MIT-Adobe FiveK Dataset [Bychkovsky et al. 2011].

### 4.4 Simulation Results

To further validate the performance improvements achieved by our proposed method, we conduct simulation experiments. Figure 11 presents the simulated holographic images alongside their corresponding color distributions on the CIE 1931 chromaticity diagram. Each pixel from the simulated image is projected onto the diagram, providing a clear visual representation of color fidelity.

As shown in Fig. 11, our method achieves the highest color fidelity among all compared approaches. In contrast, CITL+CCM yields the lowest color accuracy, exhibiting pronounced color shifts. The CITL baseline also suffers from noticeable color deviations, highlighting its limitations in preserving color fidelity. These simulation results are consistent with the optical experiments, further reinforcing the conclusion that our approach substantially improves color fidelity.

### 4.5 Ablation Study on Color Management Framework

Our proposed perceptually-aware color management framework integrates three key components: CST, MLP, and CBC. To assess the contribution of each component, we perform an ablation study in simulation. Table 1 summarizes the results across all possible component combinations integrated into the CITL baseline. Performance is evaluated using $\Delta E$, PSNR, and FSIMc metrics. The results indicate that the combination of CBC and MLP yields the most substantial improvements in image quality and color fidelity. Importantly, incorporating all three components achieves the best overall performance, underscoring the effectiveness and synergy of the proposed framework.

## 5 User Study Design and Validation

Our framework incorporates perceptually informed modeling, including MLP-based restoration to the XYZ color space to better align reconstructed images with human visual perception, evaluated using standard image quality metrics. However, such metrics may not fully capture the perceptual experience of human observers. To further validate the effectiveness of our approach, we conducted a controlled user study to assess the perceptual color fidelity of different CGH rendering algorithms. This study investigates whether participants can reliably distinguish color differences across methods, whether our framework yields statistically significant improvements, and how these differences manifest across various color categories. The following sections detail the system setup, participant recruitment, stimuli, procedure, and evaluation methodology.

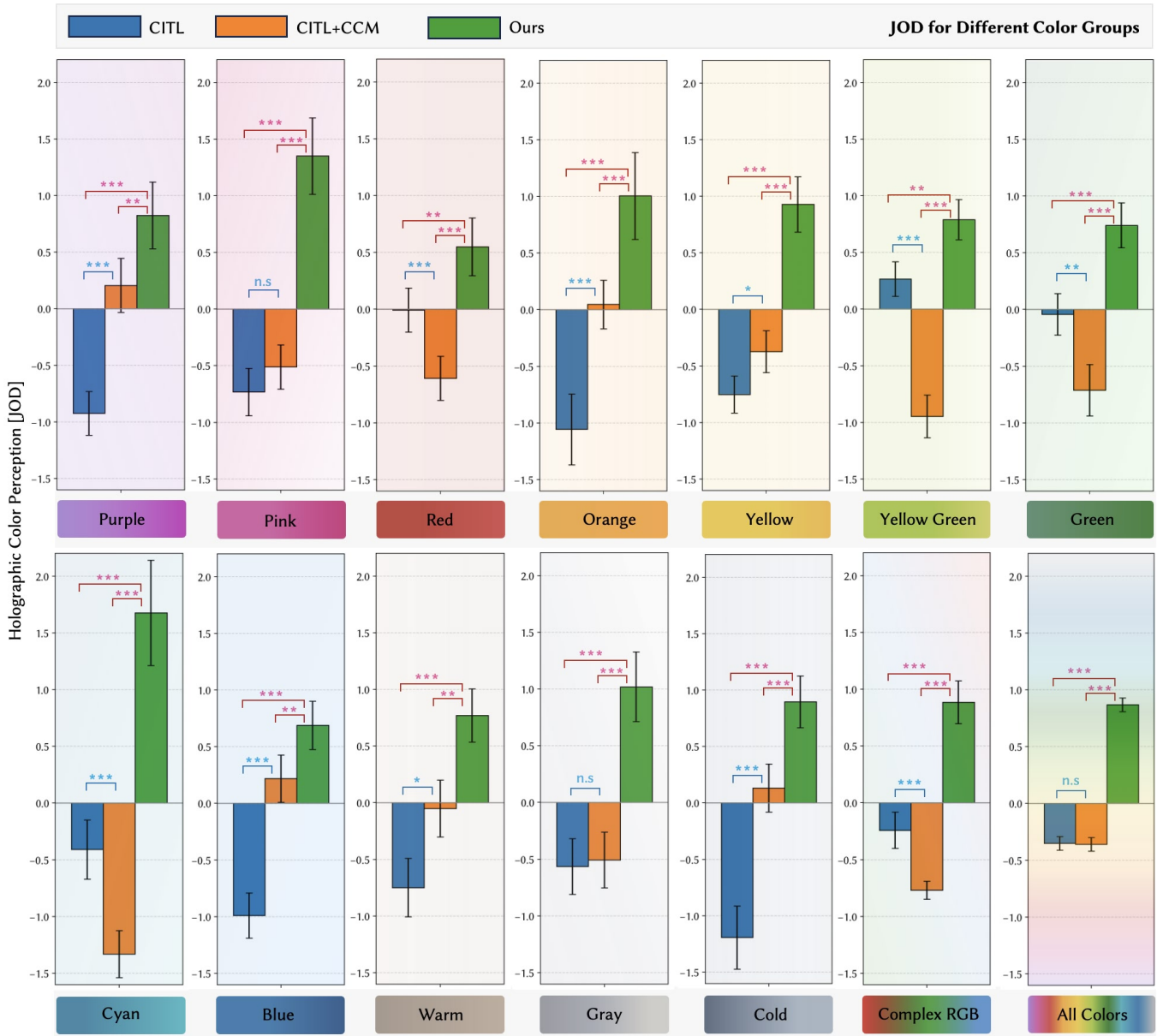

Fig. 13. User study results: Holographic color perception is evaluated for three different CGH rendering methods (CITL in blue, CITL+CCM in orange, and Ours in green) across 13 categories—including specific hue groups, color temperatures, and complex RGB scenes—as well as the overall average. The mean JOD is normalized to zero for each color group. Error bars represent 95% confidence intervals estimated from 1000 bootstrap samples. Asterisks (blue: CITL vs. CITL+CCM, red: Ours vs. competing methods) indicate statistical significance of differences (*: $p < 0.05$, **: $p < 0.01$, ***: $p < 0.001$).

*Subjects.* We recruited 20 naïve participants under the age of 40 (ranging from 23 to 38 with an average of 28.7, 15 female and 5 male), with normal or corrected-to-normal vision. The age restriction helps reduce the impact of age-related changes in visual function [Owsley 2016]. All participants passed the Ishihara color vision test and provided voluntary written informed consent. The study was approved by the Institutional Review Board of the host institution, and conducted in accordance with the Declaration of Helsinki. Participants were compensated for their time.

*Stimuli.* For the projection system, 31 scene images were first mapped to the intersection of the ProPhoto, BT.2020, and Holo color gamuts using a chroma soft-compression algorithm in the CIE $L^*C^*h$ space. Specifically, we employed a hyperbolic tangent function to compress chroma while preserving lightness and hue. The images were then converted to BT.2020 for display as ground-truth reference images. For the holographic near-eye system, we generated CGH datasets using three rendering methods: CITL, CITL+CCM, and our proposed method, across 31 scenes. To facilitate pairwise comparison between rendering methods, we prepared three sets of method-pair combinations: CITL vs. Ours, CITL vs. CITL+CCM, and Ours vs. CITL+CCM. Each combination was applied across all 31 test scenes, resulting in a total of 93 pairs used as stimuli in each session.

*User Study Configuration and Procedure.* The user study employed a dual-display setup consisting of a holographic near-eye display and a projection reference system (Fig. 12). The holographic display integrates a 1920×1080 SLM synchronized with RGB laser diodes, followed by a $4-f$ filtering system and a 40 mm focal length eyepiece that forms an image approximately 0.5 m in front of the viewer. The resulting near-eye configuration provides a 2.2 mm×2.2 mm eyebox (blue illumination) and a $12° \times 6.75°$ FOV. For ground-truth comparison, we used a commercial laser projector (XGIMI RS 10) with high color accuracy ($\Delta E < 1$) and BT.2020 gamut coverage. The projection distance was set to 0.5 m, producing a 10.5 cm×5.9 cm image. Neutral density filters were placed in front of both systems to match brightness, and their vertical positions were carefully aligned to ensure consistent viewing geometry.

Participants were seated with a chin rest, whose position was adjusted using real-time pupil data from an eye tracker mounted in front of the eyepiece. The study followed a two-interval forced-choice (2IFC) paired-comparison protocol [Bogacz et al. 2006; Chen et al. 2024b], chosen for its lower noise relative to direct rating methods [Bogacz et al. 2006; Perez-Ortiz et al. 2019]. In each trial, two holographic images—rendered with different CGH algorithms but sharing the same target—were sequentially presented to the participant's left eye for 3 seconds each, separated by a 1-second gray screen. The right eye simultaneously viewed the projector image. The room remained dark, and participants were instructed to choose the hologram whose color appeared "closer to the projector image."

Before the formal study, participants completed a brief practice session. The main experiment comprised three sessions of 93 randomized pairwise comparisons each, with approximately 20 minutes per session and a 5-minute break between sessions.

*Results.* The CGH rendering methods were evaluated for perceptual color fidelity across 31 scenes, which were categorized into 13 color groups based on color similarity. Accumulated vote counts from a total of 20 participants were normalized and scaled into Just-Objectionable-Difference (JOD) units for each color group, as illustrated in Fig. 13. A difference of 1 JOD corresponds to a perceptual threshold at which 75% of participants preferred one rendering over the other, serving as an approximate boundary of noticeable visual objection. Importantly, since an ideal optical ground truth is unattainable in current holographic displays due to physical constraints (e.g., coherent artifacts, aberrations), these scores represent relative perceptual rankings within our specific setup rather than absolute values anchored to a perfect reference. In addition, following the procedure of Pérez-Ortiz and Mantiuk [Perez-Ortiz and Mantiuk 2017], data from two participants were excluded as statistical outliers.

A two-tailed z-test was conducted to compare the scaled JOD scores between CGH rendering methods in each color group. The results indicate that our method achieved statistically significant improvements in perceptual color fidelity compared to other CGH rendering methods across all color groups. When comparing the CITL and CITL+CCM methods, the color performance and significance levels varied by group; in particular, the pink and gray groups showed no significant differences. Notably, for some colors, CITL+CCM produced higher JOD scores than CITL alone, whereas others exhibited the opposite trend. This variability is expected since a global CCM cannot reliably correct color shifts across the entire color set under narrowband illumination. Nevertheless, it may still bring certain colors closer to the ground truth, yielding localized improvements in perceptual scores. We also assessed overall color performance by computing the weighted average of JOD scores across all scenes, which confirmed that our method demonstrated substantial improvements over the competing methods.

In summary, the user study demonstrated that our proposed method consistently outperformed competing CGH rendering methods in terms of perceptual color fidelity. The significant improvements observed across a diverse set of color groups and overall scene conditions provide strong validation of the effectiveness of our approach in achieving perceptually accurate color reproduction in holographic displays.

## 6 Discussion

In this study, we investigated the important yet relatively underexplored challenge of perceptual color fidelity in holographic displays. While recent advancements have primarily focused on artifact and noise reduction, accurate perceptual color reproduction—crucial for immersive AR/VR experiences—has received comparatively less attention. We contend that color fidelity and speckle are fundamentally distinct challenges: speckle arises from wavefront coherence, whereas color errors stem from systematic spectral and radiometric factors. Even a speckle-free hologram requires rigorous color management to achieve photorealism. Thus, our framework serves as a distinct but necessary module alongside noise-reduction pipelines.

By explicitly accounting for factors contributing to color distortion—including color gamut mismatches, illumination imbalance, and camera-induced inaccuracies—our approach effectively enhances the perceptual color realism of holographic displays. We believe this work highlights the potential benefits of incorporating perceptual color considerations into holographic display pipelines, and could offer valuable insights for future investigations and practical applications.

*Hardware/Software Constraints.* While our system improves perceptual color fidelity, it remains constrained by SLM modulation limits and optical imperfections. Although our compensation algorithms mitigate these issues, residual artifacts persist. Notably, current SLM refresh rates limit the use of time-multiplexing; incorporating high-speed SLMs represents a promising direction for future enhancement [Choi et al. 2022]. Beyond hardware constraints, our current pipeline also requires several system-specific calibrations—including laser SPD measurement, Zernike-based wavefront correction, geometric alignment, and collecting training data for the MLP. These steps are typically one-time procedures, remaining valid as long as the hardware configuration remains stable. In contrast, CST, color-balance correction, and MLP architecture itself generalize well across systems, requiring only minor adjustments when illumination spectra or sensor characteristics differ. In future work, the calibration burden could be substantially reduced by incorporating learning-based system identification: by collecting pairs of input holograms and captured outputs, data-driven neural holography models [Gopakumar et al. 2024; Peng et al. 2020; Schiffers et al. 2025] may enable automatic estimation of optical aberrations, spectral distortions, and color-correction mappings. Such an end-to-end formulation not only streamlines the calibration workflow but also provides a path toward achieving real-time hologram generation.

*User Study Challenges.* Our user study employed a dataset of 31 distinct color scenes spanning a diverse range of hues. While this dataset enabled substantial initial insights, it remains relatively coarse in granularity. Subdividing the color wheel into more refined groups could support more precise and nuanced analysis of color perception [He et al. 2023]. Recognizing practical constraints such as experimental duration and participant fatigue, future studies should aim to strike an effective balance between dataset size and experimental complexity. In addition, while the binocular comparative method allowed efficient side-by-side evaluation, it introduces potential perceptual confounds. Specifically, the dichoptic setup—where each eye views a different display—may induce binocular rivalry [Tong et al. 2006] or differential chromatic adaptation [Fairchild 2013]. Since the two eyes are exposed to different spectral power distributions, their adaptation states may diverge, potentially influencing color appearance judgments. Although we mitigated this by matching luminance, utilizing neutral gray intervals and employing short stimulus durations, these physiological factors remain a limitation of dual-display assessments. To address this, a promising direction for future studies is to integrate a laser projector into a monocular comparative holographic near-eye display system, thereby simplifying participant tasks and minimizing perceptual interference. This approach, however, poses new optical challenges, such as additional requirements for precise color calibration and chromatic aberration correction, which would require careful optical system design. Complementing these system-level corrections, incorporating user-specific ocular wavefronts [Kim et al. 2021] would further ensure fair comparisons by matching the visual sharpness of the hologram to that of the reference image, thereby minimizing the impact of blur on color judgment.

*Colorimetric Matching Failure.* Accurate CMFs are fundamental for ensuring perceptual color fidelity, as they directly influence color perception based on spectral data. The widely adopted CIE standard CMFs, including CIE 1931, CIE 1964, and CIE 2006, formed the basis of our current color management framework. However, while standard CMFs are effective for conventional broadband displays, they typically yield systematic perceptual deviations when applied to highly monochromatic laser-based systems. This is primarily due to the narrowband emission characteristics of lasers [Ko et al. 2023]. This limitation underscores an essential direction for further theoretical refinement and empirical investigation into optimized CMFs for holographic displays.

## 7 Conclusion

We present a perceptually-aware color management framework that improves the accuracy of color reproduction in holographic displays through perceptually informed optimization. By combining color space transformation, learning-based camera color restoration, and adaptive color balance with a compensated light propagation model, our method enables perceptually guided hologram rendering. Validated through simulation, optical experiments, and a controlled user study, our color management framework demonstrates significant and consistent improvements in perceptual color fidelity across diverse scenes and color conditions. We believe this work offers not only a practical solution for improving perceptual color fidelity in near-eye holographic displays, but also lays a foundation for future developments in perceptually informed color management for holographic display systems.

## Acknowledgments

This work was supported by the BK21 FOUR program of the Education and Research Program for Future ICT Pioneers, Seoul National University in 2025 [20 %]; by the Institute of Information & Communications Technology Planning & Evaluation (IITP) grant funded by the Korean Government (MSIT) (No. 2019-0-00001, Development of Holo-TV Core Technologies for Hologram Media Services) [20%]; and by the National Research Foundation of Korea (NRF) grants funded by the Korean government (MSIT) (Grant No. RS-2022-NR070432 [20%], Grant No. RS-2024-00416272 [20%], and Grant No. RS-2024-00414230 [20%]).

# PAColorHolo: A Perceptually-Aware Color Management Framework for Holographic Displays - Supplementary Material

CHUN CHEN, MINSEOK CHAE, and SEUNG-WOO NAM, Seoul National University, Republic of Korea
MYEONG-HO CHOI, MINSEONG KIM, and EUNBI LEE, Seoul National University, Republic of Korea
YOONCHAN JEONG* and JAE-HYEUNG PARK*, Seoul National University, Republic of Korea

This is a supplementary material for "PAColorHolo: A Perceptually-Aware Color Management Framework for Holographic Displays".

## S1 Implementation

### S1.1 Holographic Display System

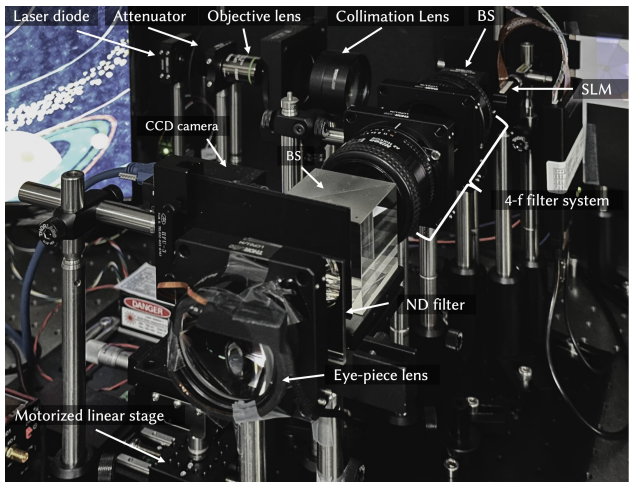

Fig. S1. Holographic display system.

The optical configuration of the proposed holographic display system is illustrated in Fig. S1. We employ an RGB fiber-coupled laser module (FISBA READYBeam) as the coherent light source. The center wavelengths of the red, green, and blue laser channels are measured to be 636 nm, 512 nm, and 453 nm, respectively, using a calibrated spectrometer. For spatial light modulation, we use a HOLOEYE LETO phase-only liquid crystal-on-silicon (LCoS) spatial light modulator (SLM), featuring a resolution of $1920 \times 1080$ pixels and a pixel pitch of 6.4 μm. The SLM supports 8-bit phase modulation and achieves a diffraction efficiency exceeding 93%. The SLM operates in a color-field-sequential (CFS) mode, synchronized with the RGB laser channels via the Arduino MEGA microcontroller. This configuration enables full-color holographic display at a frame rate of 60 Hz. A standard $4-f$ optical setup is implemented to suppress higher-order diffraction orders and noise. For image acquisition, we use a color CCD camera (Grasshopper3 GS3-U3-89S6C) operating in raw mode. The captured raw images undergo only demosaicing to reconstruct RGB values, preserving the original sensor data without additional processing. The exposure time is set to $n/60$ seconds, where $n$ is an integer, synchronized with the display frame cycle. The camera is mounted on a motorized linear translation stage, allowing precise image capture at varying depths to support focal stack acquisition. To deliver the reconstructed holographic image to the eye, a 45 mm focal length eyepiece lens is employed. A neutral density (ND) reflective filter is inserted before the eyepiece to regulate perceived brightness, ensuring both comfortable and safe viewing conditions.

### S1.2 User Study System

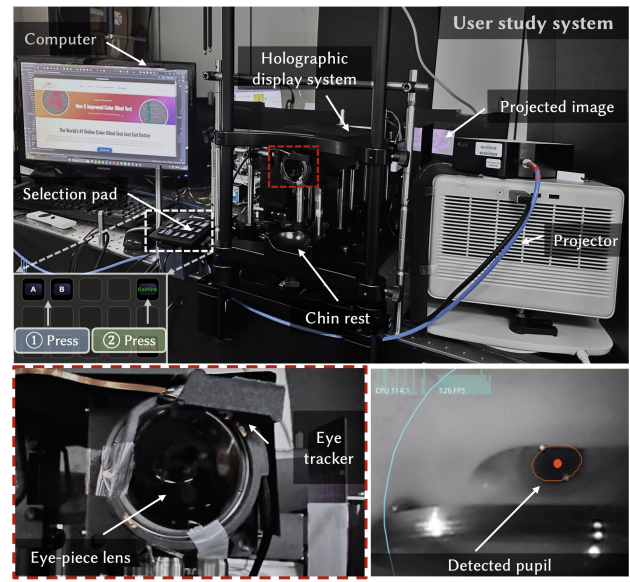

Fig. S2. User study system.

The user study system is illustrated in Fig. S2, comprising a holographic near-eye display and a laser projection display. An integrated eye tracker detects the participant's pupil position, and the chin rest is adjusted accordingly to ensure proper alignment with the holographic display system. The participant's left and right eyes simultaneously view the holographic and projected images, enabling direct binocular color comparison.

*Corresponding authors

Authors' Contact Information: Chun Chen, chenchun@snu.ac.kr; Minseok Chae, mschae3d@gmail.comm; Seung-Woo Nam, 711asd@snu.ac.kr, Seoul National University, Seoul, Republic of Korea; Myeong-Ho Choi, mhchoi960905@snu.ac.kr; Minseong Kim, vkminseong@snu.ac.kr; Eunbi Lee, dldmsql0215@snu.ac.kr, Seoul National University, Seoul, Republic of Korea; Yoonchan Jeong, yoonchan@snu.ac.kr; Jae-Hyeung Park, jaehyeung@snu.ac.kr, Seoul National University, Seoul, Republic of Korea.

Participants made their selections via a dedicated input pad. During each comparison trial, a verbal prompt was played to indicate the order of Image A and Image B and to cue a response. Participants responded by pressing either the "A" or "B" button, followed by a "Confirm" button to finalize their choice. Each comparison lasted approximately 10 seconds. All response data were automatically recorded and saved as CSV files on the host computer.

## S2 Algorithmic Procedures and Pseudocode

This section provides pseudocode for all algorithms presented in the main paper. Algorithm S1 outlines the complete optimization procedure of the proposed PAColorHolo method (Section 3.5). Algorithm S2 details the training process of the multilayer perceptron (MLP) based camera color restoration model (Section 3.3). Algorithm S3 describes the optimization procedure of the baseline camera-in-the-loop (CITL) method. Lastly, Algorithms S4 and S5 present the color correction matrix (CCM) optimization using a color checker and the integration of CCM into the CITL+CCM rendering pipeline, respectively (Section 4.3).

**Algorithm S1** PAColorHolo Optimization Process

1: $R(u, v)$ : replace values of $u$ with $v$ while retaining gradients from $u$ for backpropagation
2: $I_c, I_o$ : captured and target images in their original color spaces
3: $I_{cholo}, I_{tholo}$ : captured and target images in the Holo color space
4: $\mathcal{P}(\phi)$ : light propagation model with input phase $\phi$
5: $s$ : learnable energy scaling factor
6: $\alpha_\phi, \alpha_s$ : learning rates for phase and scaling factor updates
7: $K$ : number of optimization iterations
8: $\mathcal{L}$ : mean squared error (MSE) loss function
9: $CBC\{\cdot\}$ : color balance correction
10: $CST\{\cdot\}$ : color space transformation
11: $MLP\{\cdot\}$ : color restoration with the trained MLP
12: **Input:** $I_{tholo} = CST\{I_o; C_o, C_{holo}\}$ ▷ to Holo color space
13: **Initialize:** Phase profile $\phi \leftarrow \phi_0$ ▷ random initialization
   Scaling factor $s \leftarrow s_0 = 1$ ▷ constant initialization
   $r, g, b \leftarrow CBC\{CST\{I_{tholo}; C_{holo}, C_{XYZ}\}\}$ ▷ laser ini.
14: **for** $k=1$ to $K$ **do**
15: $\quad I_{cholo} = CST\{MLP\{I_c(\phi)\}; C_{XYZ}, C_{holo}\}$ ▷ to Holo color space
16: $\quad \phi \leftarrow \phi - \alpha_\phi \nabla_\phi \left( \mathcal{L}\left(s \cdot R(|\mathcal{P}(\phi)|^2, I_{cholo}), I_{tholo}\right) \right)$
17: $\quad s \leftarrow s - \alpha_s \nabla_s \left( \mathcal{L}\left(s \cdot R(|\mathcal{P}(\phi)|^2, I_{cholo}), I_{tholo}\right) \right)$
18: $\quad$ **if** $k = K/2$ **then**
19: $\quad\quad r, g, b \leftarrow CBC\{CST\{I_{cholo}; C_{holo}, C_{XYZ}\}\}$ ▷ laser update
20: $\quad$ **end if**
21: **end for**
22: **return** $\phi, s$

**Algorithm S2** MLP Training for Color Restoration

1: $\mathcal{D}_m = \{(x_i^m, y_i^m)\}, \mathcal{D}_c = \{(x_i^c, y_i^c)\}$ : monochromatic and polychromatic training datasets (camera RGB → target XYZ)
2: $\mathcal{M}_\theta$ : MLP model with parameters $\theta$
3: $\mathcal{L}$ : L1 loss function
4: $\eta$ : learning rate
5: $N$ : number of training epochs
6: $\Delta E$ : color difference between MLP prediction and ground truth
7: **Input:** Datasets $\mathcal{D}_m$ and $\mathcal{D}_c$
8: **Initialize:** $\theta \leftarrow \theta_0$ ▷ random initialization
9: **for** $epoch$ = 1 to $N$ **do**
10: $\quad$ **for** each iteration **do**
11: $\quad\quad (x^m, y^m) \sim \mathcal{D}_m, \ (x^c, y^c) \sim \mathcal{D}_c$
12: $\quad\quad \hat{y}^m \leftarrow \mathcal{M}_\theta(x^m), \quad \hat{y}^c \leftarrow \mathcal{M}_\theta(x^c)$
13: $\quad\quad \mathcal{L}_{train} \leftarrow \mathcal{L}(\hat{y}^m, y^m) + \mathcal{L}(\hat{y}^c, y^c)$
14: $\quad\quad \theta \leftarrow \theta - \eta \cdot \nabla_\theta \mathcal{L}_{\text{train}}$
15: $\quad$ **end for**
16: $\quad$ Evaluate on validation set: report average L1 loss and $\Delta E$
17: $\quad$ **if** validation loss improves **then**
18: $\quad\quad$ Save model parameters $\theta$
19: $\quad$ **end if**
20: **end for**
21: **return** trained model $\mathcal{M}_\theta$

**Algorithm S3** CITL Optimization Process

1: $R(u, v)$ : replace values of $u$ with $v$ while retaining gradients from $u$ for backpropagation
2: $I_c, I_o$ : captured and target images in their original color spaces
3: $I_{tholo}$ : target image in the holographic color space
4: $\mathcal{P}(\phi)$ : light propagation model with input phase $\phi$
5: $s$ : learnable energy scaling factor
6: $\alpha_\phi, \alpha_s$ : learning rates for phase and scaling factor updates
7: $K$ : number of optimization iterations
8: $\mathcal{L}$ : MSE loss function
9: **Input:** $I_o$
10: **Initialize:** Phase profile $\phi \leftarrow \phi_0$ ▷ random initialization
    scaling factor $s \leftarrow s_0 = 1$ ▷ constant initialization
    $r, g, b \leftarrow CBC\{CST\{I_{tholo}; C_{holo}, C_{XYZ}\}\}$ ▷ laser ini.
11: **for** $k=1$ to $K$ **do**
12: $\quad \phi \leftarrow \phi - \alpha_\phi \nabla_\phi \left( \mathcal{L}\left(s \cdot R(|\mathcal{P}(\phi)|^2, I_c(\phi)), I_o\right) \right)$
13: $\quad s \leftarrow s - \alpha_s \nabla_s \left( \mathcal{L}\left(s \cdot R(|\mathcal{P}(\phi)|^2, I_c(\phi)), I_o\right) \right)$
14: **end for**
15: **return** $\phi, s$

**Algorithm S4** CCM Optimization

1: $X_{RGB}, Y_{RGB}$ : captured linear RGBs and ground truth linear RGBs from the color checker
2: $W = \text{diag}(w_1, w_2, w_3)$ : learnable diagonal white balance matrix
3: $M$ : learnable color correction matrix
4: $M_c$ : overall color correction matrix combining color correction and white balancing
5: $K_1, K_2$ : number of optimization iterations for WB and CCM
6: $\eta_W, \eta_M$ : learning rates for $W$ and $M$ updates
7: $\mathcal{L}$ : loss function (e.g., MSE or patch-based error)
8: $\mathcal{WB}\{X, W\}$ : apply white balance matrix to RGB data
9: $CC\{X, M\}$ : apply CCM to RGB data
10: **Input:** $X_{RGB}, Y_{RGB}$
11: **Initialize:** $W \leftarrow W_0, M \leftarrow M_0$ ▷ random initialization
12: **for** $t = 1$ to $K_1$ **do** ▷ White balance matrix optimization
13:     $X' = \mathcal{WB}\{X_{RGB}, W\}$
14:     $W \leftarrow W - \eta_W \cdot \nabla_W \mathcal{L}(X', Y_{RGB})$
15: **end for**
16: $X' \leftarrow \mathcal{WB}\{X_{RGB}, W\}$
17: **for** $t = 1$ to $K_2$ **do** ▷ Color correction matrix optimization
18:     $\hat{Y} = CC\{X', M\}$
19:     $M \leftarrow M - \eta_M \cdot \nabla_M \mathcal{L}(\hat{Y}, Y_{RGB})$
20: **end for**
21: **return** $M_c = M \cdot W$

**Algorithm S5** CITL+CCM Optimization

1: $R(u, v)$ : replace values of $u$ with $v$ while retaining gradients from $u$ for backpropagation
2: $I_c, I_o$ : captured and target images in their original color spaces
3: $I_{tholo}$ : target image in the holographic color space
4: $\mathcal{P}(\phi)$ : light propagation model with input phase $\phi$
5: $s$ : learnable energy scaling factor
6: $\alpha_\phi, \alpha_s$ : learning rates for phase and scaling factor updates
7: $K$ : number of optimization iterations
8: $\mathcal{L}$ : MSE loss function
9: $CC\{I, CCM\}$: Apply CCM to image $I$
10: **Input:** $I_{tholo} = CST\{I_o; C_o, C_{holo}\}$ ▷ to holo color space
11: **Initialize:** Phase profile $\phi \leftarrow \phi_0$ ▷ random initialization
    scaling factor $s \leftarrow s_0 = 1$ ▷ constant initialization
    $r, g, b \leftarrow CBC\{CST\{I_{tholo}; C_{holo}, C_{XYZ}\}\}$ ▷ laser ini.
12: **for** $k=1$ to $K$ **do**
13:     $I_{CCM} = CC\{I_c(\phi), M_c\}$
14:     $\phi \leftarrow \phi - \alpha_\phi \nabla_\phi \left( \mathcal{L}\left(s \cdot R(|\mathcal{P}(\phi)|^2, I_{CCM}), I_o\right) \right)$
15:     $s \leftarrow s - \alpha_s \nabla_s \left( \mathcal{L}\left(s \cdot R(|\mathcal{P}(\phi)|^2, I_{CCM}), I_o\right) \right)$
16: **end for**
17: **return** $\phi, s$

## S3 Color Space Definition and Transformation

### S3.1 Definition of Holographic Color Space

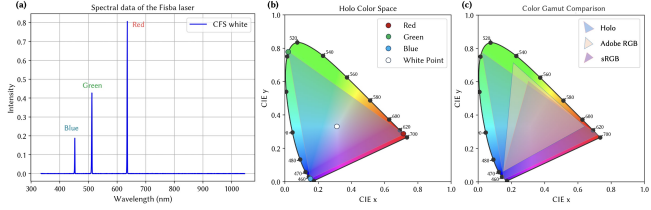

Fig. S3. (a) Experimentally measured SPDs of the RGB laser sources. (b) The defined Holo color space visualized on the CIE 1931 chromaticity diagram, constructed based on the measured primaries and corresponding white point. (c) Color gamut comparison among Holo, Adobe RGB (1998) and sRGB color spaces.

The holographic color space (Holo color space) gamut border can be defined by specifying primary colors derived from the spectral power distributions (SPDs) of the lasers used in holographic displays. Figure S3(a) shows the experimentally measured normalized SPD of our laser. Given the standard CIE 1931 color matching functions (CMFs), the tristimulus values (X, Y, Z) of each primary color can be obtained by discretely integrating over the spectral distribution as follows:

$$X = k \sum_{i=1}^{N} P_i \, \bar{x}_i \, \Delta\lambda, \tag{1}$$

$$Y = k \sum_{i=1}^{N} P_i \, \bar{y}_i \, \Delta\lambda, \tag{2}$$

$$Z = k \sum_{i=1}^{N} P_i \, \bar{z}_i \, \Delta\lambda, \tag{3}$$

where $P_i$ denotes the sampled SPD at wavelength $\lambda_i$, $\bar{x}_i$, $\bar{y}_i$, and $\bar{z}_i$ are the CIE 1931 color matching functions at wavelength $\lambda_i$, and $\Delta\lambda$ is the sampling interval. The constant $k$ normalizes the tristimulus values, typically set to $k = 1$ for relative calculations.

From these tristimulus values, the chromaticity coordinates $(x, y)$ for each primary are computed:

$$x = \frac{X}{X+Y+Z}, \quad y = \frac{Y}{X+Y+Z}. \tag{4}$$

We adopt the D65 standard illuminant as the white point for the holographic color space, due to its widespread usage in the display industry. It is represented as $w = [X_w, Y_w, Z_w]^T$, corresponding to the standard CIE tristimulus values of D65.

Consequently, the holographic color space ($C_{holo}$) is explicitly defined by its three primaries and the white point:

$$C_{holo} = \{\mathbf{r}, \mathbf{g}, \mathbf{b}, \mathbf{w}\}. \tag{5}$$

The defined Holo color space is visualized in the CIE chromaticity diagram, as illustrated in Fig. S3(b). To further explain the practical implications of the Holo color space, Fig. S3(c) presents a direct comparison between the Holo color space and standard color spaces (sRGB and Adobe RGB). As observed in Fig. S3(c), a distinct characteristic of the Holo color space is the position of its primaries. Unlike standard RGB color spaces whose primaries are specified within the spectral locus, the laser primaries are much closer to the spectral locus due to their narrowband SPDs. This visual comparison highlights the significant gamut mismatch between the standard color spaces and the holographic display.

### S3.2 Color Space Transformation

To enable practical conversion between the Holo color space and standard color spaces, we define a linear Color Space Transform (CST). Given a source color space $C_s$ and a target color space $C_t$, we construct forward and inverse transformation matrices based on the chromaticity coordinates of the primaries and their corresponding white points.

Let the $3 \times 3$ matrix $\mathbf{M}$ represent the unscaled linear RGB-to-XYZ transformation matrix for a given color space, formed from the XYZ tristimulus values of the red, green, and blue primaries:

$$\mathbf{M} = \begin{bmatrix} X_r & X_g & X_b \\ Y_r & Y_g & Y_b \\ Z_r & Z_g & Z_b \end{bmatrix}. \tag{6}$$

To ensure that white maps to the correct reference white point $w$, we solve for a diagonal scaling vector:

$$\mathbf{S} = \mathbf{M}^{-1} \begin{bmatrix} X_w \\ Y_w \\ Z_w \end{bmatrix}. \tag{7}$$

The calibrated RGB-to-XYZ transformation matrix is then given by:

$$\mathbf{M}_{\text{RGB}\to\text{XYZ}} = \mathbf{M} \cdot \text{diag}(S_r, S_g, S_b), \tag{8}$$

and its inverse (XYZ to RGB) is:

$$\mathbf{M}_{\text{XYZ}\to\text{RGB}} = \mathbf{M}_{\text{RGB}\to\text{XYZ}}^{-1}. \tag{9}$$

The general linear CST operator from a source color space $C_s$ to a target color space $C_t$ is therefore:

$$I_t = CST\{I_s; C_s, C_t\} = \mathbf{M}_{\text{XYZ}\to C_t} \, \mathbf{A}_{w_s \to w_t} \, \mathbf{M}_{C_s \to \text{XYZ}} \, I_s, \tag{10}$$

where $I_s$ and $I_t$ denote the RGB values of the image in the source and target color spaces, and $\mathbf{A}_{w_s \to w_t}$ is an optional chromatic adaptation matrix, which becomes the identity matrix when the white points of the two color spaces are the same. Similarly, when either $C_s$ or $C_t$ corresponds to the CIE XYZ color space, the associated transformation matrix $\mathbf{M}$ also becomes an identity matrix.

## S4 Optical Imperfections in Holographic Display Systems

Holographic display systems inherently suffer from optical and chromatic aberrations introduced by cascaded optical components. These system-induced imperfections can significantly degrade the fidelity of the reconstructed image. As a result, effective aberration modeling and compensation techniques are critical for achieving high-quality holographic reproduction.

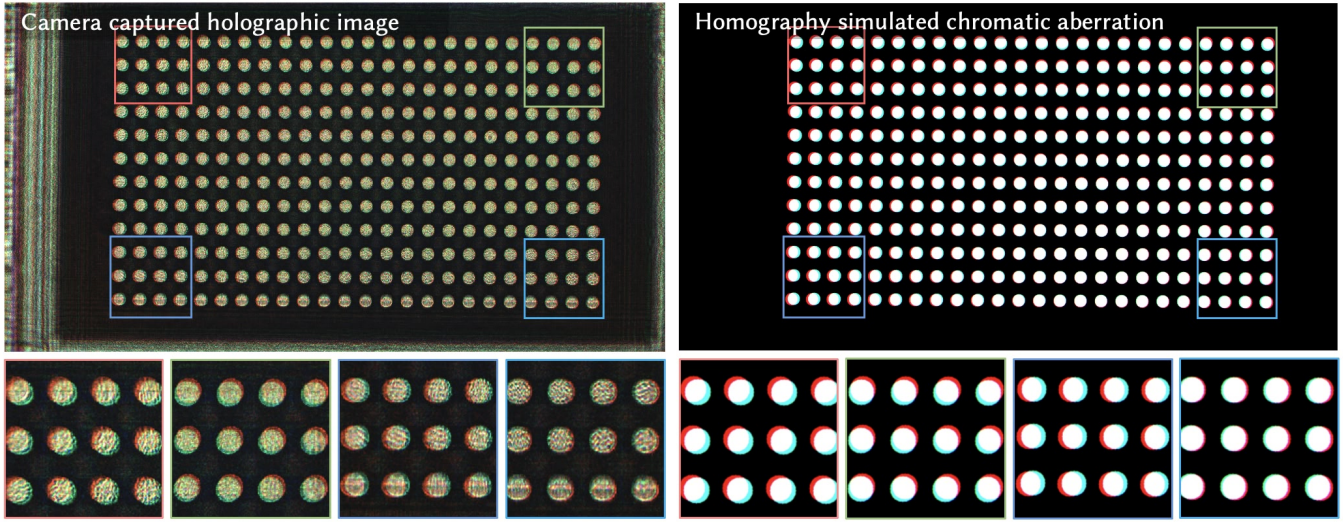

Fig. S4. Chromatic aberration modeling. The left image displays a captured circle array with pronounced chromatic aberration, while the right image shows the simulated distortion produced using inverse homography, demonstrating high visual consistency with the experimental result.

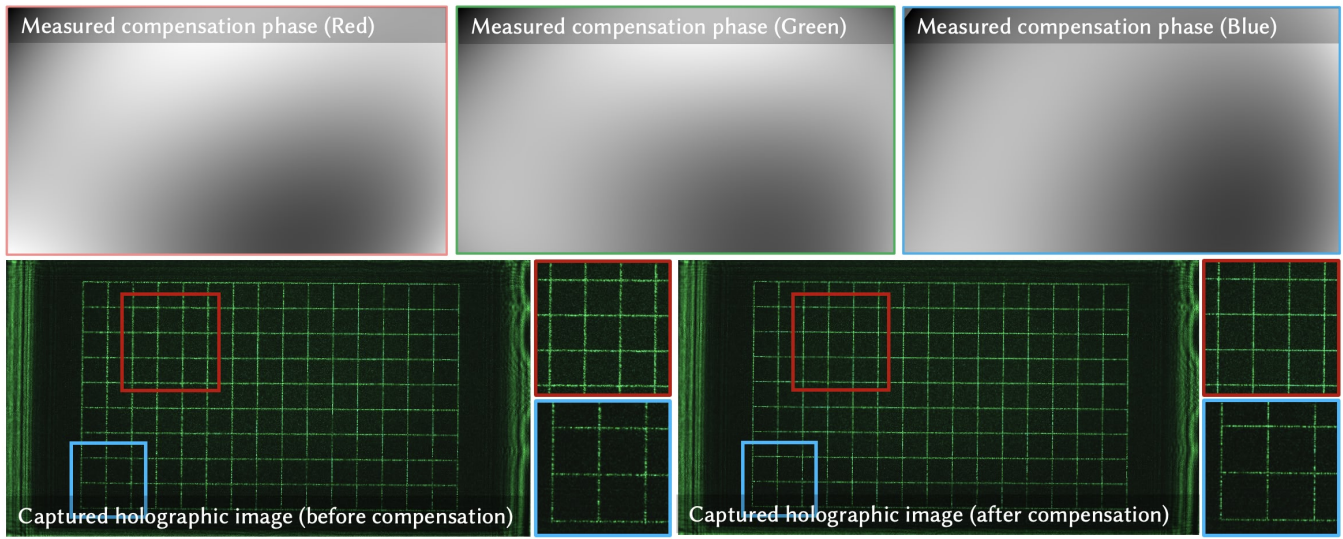

Fig. S5. Optical aberration compensation. The first row shows the experimentally estimated compensation phase represented using Zernike polynomials. The second row compares reconstructed holographic images without aberration compensation (left) and with compensation (right), demonstrating improved image quality.

### S4.1 Homography for Chromatic Aberration Correction

Chromatic aberrations in holographic optical systems are often complex and cannot be effectively corrected by simply scaling the RGB channels of the target image. To model these aberrations, we employ an inverse homography matrix that approximates the wavelength-dependent spatial distortions observed in the numerically reconstructed image. The homography matrix generation follows the CITL method [Peng et al. 2020].

Once modeled, the chromatic aberration can be compensated through the hologram optimization process. As illustrated in Fig. S4, the left panel shows a camera-captured holographic image with pronounced chromatic aberration, while the right panel presents the simulated distortion using the inverse homography. The strong alignment between the two validates the accuracy of the simulation and supports its use in compensation.

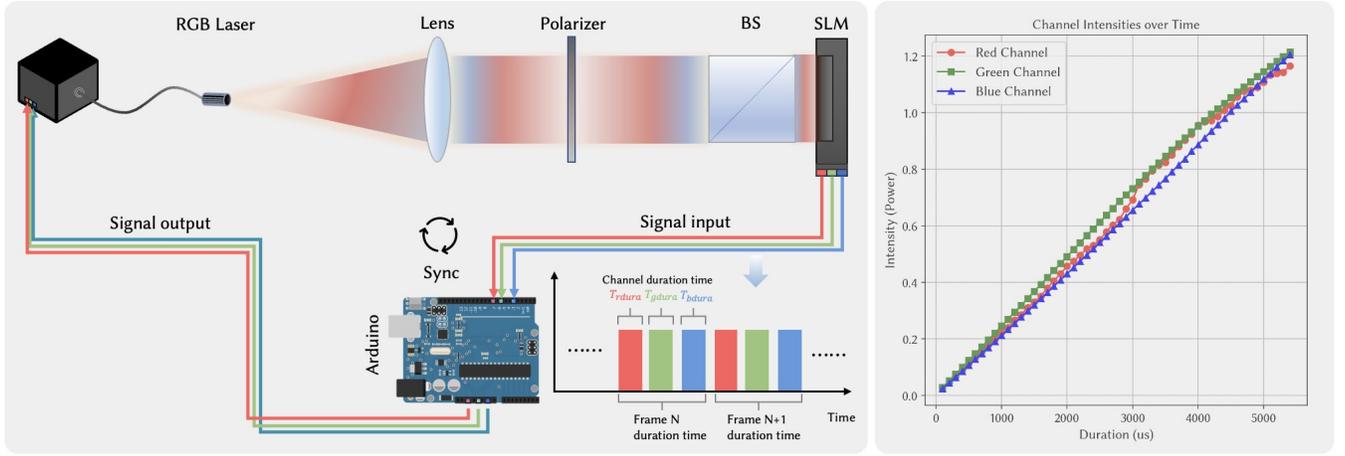

Fig. S6. Full-color holographic display enabled by synchronized operation of the SLM and RGB lasers using an Arduino microcontroller (left). Measured intensity of each RGB channel as a function of illumination duration (right), captured using a power meter, demonstrates the approximately linear behavior essential for color balance control.

### S4.2 Aberration Correction with Zernike Compensation Phase

To account for optical aberrations in the system, we incorporate a Zernike compensation phase into the propagation model. The compensation phase $\phi_c$ is represented as a linear combination of Zernike polynomials [Kim et al. 2021]:

$$\phi_c(\rho, \theta) = \sum_{j=1}^{J} c_j Z_j(\rho, \theta). \tag{11}$$

Here, $\rho$ and $\theta$ represent the normalized radial and angular coordinates, respectively; $Z_j$ denotes the $j$-th Zernike polynomial following Noll's sequential indexing, and $c_j$ are the corresponding coefficients. These coefficients are experimentally estimated to characterize the system's aberrations. Since the SLM has a rectangular aperture, we generate the Zernike compensation phase by first computing a circular phase map defined over the circumscribed circle of the SLM's active area, and then cropping it to fit the rectangular display region. This ensures compatibility between the circular Zernike model and the physical geometry of the SLM. The resulting compensation phases for the RGB channels are shown in the first row of Fig. S5. The second row presents the reconstructed holographic images without and with aberration compensation, highlighting the effectiveness of the correction.

### S5 Color Balance Correction Details

#### S5.1 Color-field-sequential model of the SLM

The CFS operation mode of the SLM plays a critical role in enabling perceptually accurate color balance correction in our holographic display system. In CFS mode, each RGB channel is illuminated sequentially, and the resulting intensity exhibits an approximately linear relationship with the illumination duration, as shown in Fig. S6. This near-linear behavior provides a reliable basis for modulating RGB power ratios and facilitates the implementation of our color balance correction (CBC) strategy.

In addition, CFS operation is essential for the user study, as it allows the system to display full-color holographic images in real time. To support this functionality, we synchronize the SLM with the RGB laser sequence using an Arduino controller, enabling stable full-color display at 60 Hz.

#### S5.2 Laser Intensity to Color

In the CFS model, the relationship between the laser output intensity and the duration of each channel is approximately linear. Thus, we estimate the laser intensity for each RGB channel as:

$$\begin{aligned} R_{\text{int}} &= R_{\max} \cdot T_r, \\ G_{\text{int}} &= G_{\max} \cdot T_g, \\ B_{\text{int}} &= B_{\max} \cdot T_b, \end{aligned} \tag{12}$$

where $R_{\max}, G_{\max}, B_{\max}$ denote the measured intensities of the RGB laser channels at a reference exposure duration (e.g., 4 ms), and $T_r, T_g, T_b$ are the programmed durations for each channel.

Given the narrowband nature of laser diodes, the RGB intensities can be treated as a discrete spectral power distribution. To convert these into colorimetric (XYZ) values, we utilize the CIE 1931 CMFs evaluated at the central wavelengths of the RGB lasers. This yields a $3 \times 3$ matrix $M_{\text{CMFs}}$:

$$M_{\text{CMFs}} = \begin{bmatrix} \bar{x}(\lambda_R) & \bar{x}(\lambda_G) & \bar{x}(\lambda_B) \\ \bar{y}(\lambda_R) & \bar{y}(\lambda_G) & \bar{y}(\lambda_B) \\ \bar{z}(\lambda_R) & \bar{z}(\lambda_G) & \bar{z}(\lambda_B) \end{bmatrix}, \tag{13}$$

where $\bar{x}(\cdot), \bar{y}(\cdot), \bar{z}(\cdot)$ are the CIE 1931 CMFs, and $\lambda_R, \lambda_G, \lambda_B$ are the central wavelengths of the RGB lasers.

Using this matrix, we convert the laser intensity vector to CIE XYZ tristimulus values:

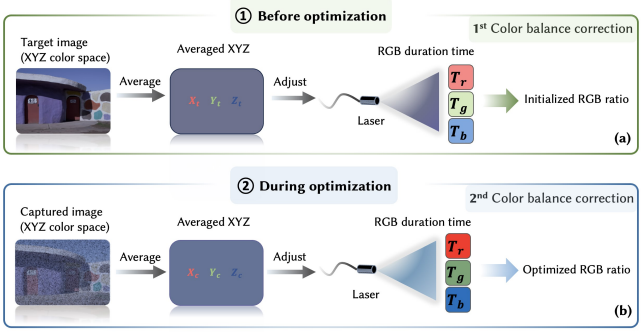

Fig. S7. Two-step color balance correction for the holographic display. (a) At the initial stage of optimization, the laser's RGB power ratio is computed based on the target image color. (b) During the intermediate stage of optimization, the RGB laser duration is adaptively updated based on the captured image color (processed via $MLP\{\cdot\}$) to enhance perceptual fidelity. Images Credits: MIT-Adobe FiveK Dataset [Bychkovsky et al. 2011].

$$\begin{bmatrix} X_{\text{int}} \\ Y_{\text{int}} \\ Z_{\text{int}} \end{bmatrix} = M_{\text{CMFs}} \cdot \begin{bmatrix} R_{\text{int}} \\ G_{\text{int}} \\ B_{\text{int}} \end{bmatrix}. \tag{14}$$

### S5.3 Two-Step Color Balance Correction

We present a two-step CBC method to enhance perceptual color fidelity in holographic displays. As illustrated in Fig. S7, the top row shows the first-stage correction, which derives the laser RGB ratio from the target image. The bottom row depicts the second-stage correction, which refines the RGB ratio based on the color distribution of the captured image. This method exploits the near-linear relationship between laser intensity and exposure duration under the CFS model, and uses the CIE 1931 color matching functions to convert laser spectral power distributions into CIE XYZ tristimulus estimates for colorimetric color balancing, improving perceptual color fidelity.

**Step 1: Initial RGB Duration Estimation.** To match the overall color of the reconstructed holographic image with the target, we first convert the average XYZ value of the target image into laser intensity using the inverse of the CMF matrix:

$$[R_{\text{int}}, G_{\text{int}}, B_{\text{int}}] = M_{\text{CMFs}}^{-1} \cdot \begin{bmatrix} X_{\text{tar}} \\ Y_{\text{tar}} \\ Z_{\text{tar}} \end{bmatrix} \tag{15}$$

Given the measured peak intensities $[R_{\max}, G_{\max}, B_{\max}]$, the initial RGB duration times are computed via element-wise division:

$$[T_r^{1\text{st}}, T_g^{1\text{st}}, T_b^{1\text{st}}] = \frac{[R_{\text{int}}, G_{\text{int}}, B_{\text{int}}]}{[R_{\max}, G_{\max}, B_{\max}]} \tag{16}$$

**Step 2: Adaptive RGB Refinement.** As the hologram optimization redistributes energy among RGB channels, a single initialization is insufficient. After several iterations, we capture the rendered holographic image and apply our trained color restoring multilayer perceptron (MLP) neural-network to obtain the corresponding average XYZ values:

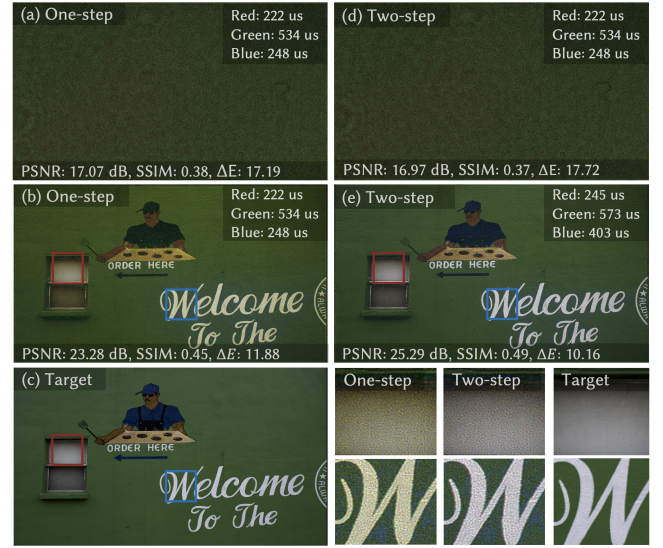

Fig. S8. Comparison of holographically reconstructed images using (a)-(b) one-step and (d)-(e) two-step color balance correction. The corresponding RGB laser durations are displayed. (c)The target image. Images Credits: MIT-Adobe FiveK Dataset [Bychkovsky et al. 2011].

$$[X_{\text{cap}}, Y_{\text{cap}}, Z_{\text{cap}}] = MLP\{I_{\text{cap}}\} \tag{17}$$

The captured RGB intensities are then recovered as:

$$[R_c, G_c, B_c] = M_{\text{CMFs}}^{-1} \cdot \begin{bmatrix} X_{\text{cap}} \\ Y_{\text{cap}} \\ Z_{\text{cap}} \end{bmatrix} \tag{18}$$

The intensity correction ratios between captured and initial values are computed via element-wise division:

$$[\text{Ratio}_r, \text{Ratio}_g, \text{Ratio}_b] = \frac{[R_c, G_c, B_c]}{[R_{\text{int}}, G_{\text{int}}, B_{\text{int}}]} \tag{19}$$

The updated RGB durations are computed via element-wise division:

$$[T_r^{2\text{nd}}, T_g^{2\text{nd}}, T_b^{2\text{nd}}] = \frac{[T_r^{1\text{st}}, T_g^{1\text{st}}, T_b^{1\text{st}}]}{[\text{Ratio}_r, \text{Ratio}_g, \text{Ratio}_b]} \tag{20}$$

Figures S8(a)–(b) show holographic reconstruction results using the first-step CBC. While this initial RGB ratio effectively preserves dominant colors, it often leads to noticeable color shifts—particularly in neutral regions such as white—due to subsequent hologram optimization steps that redistribute RGB energy. Figures S8(d)–(e) present the results after applying the second-step CBC. This additional adjustment significantly mitigates color shifts, resulting in overall color tones that more closely match the target image shown in Fig. S8(c).

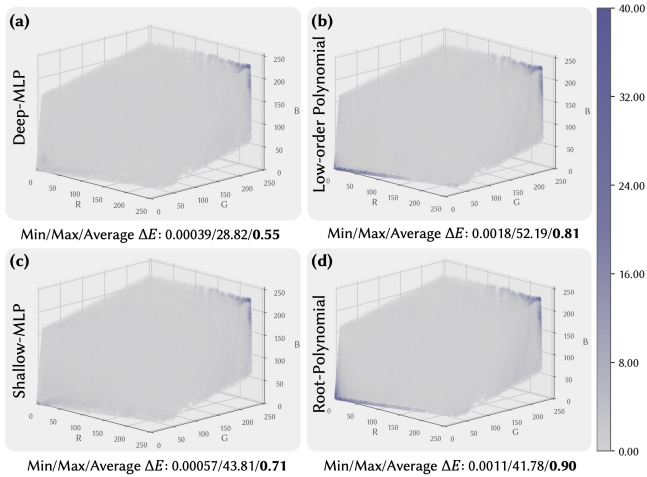

Fig. S9. ΔE performance comparison among different color mapping methods: (a) the proposed deep MLP, (b) low-order polynomial, (c) shallow MLP, and (d) root-polynomial. The deep MLP demonstrates superior performance (lowest ΔE), particularly in low-luminance regions where polynomial methods underperform.

## S6 Camera Calibration

### S6.1 Camera's Dark Field Correction

To compensate for the fixed sensor bias in the camera output, we apply a dark field correction step prior to further processing. Specifically, a constant offset of 15 (on an 8-bit scale) is subtracted from all pixel values in the raw 8-bit images. This value approximates the baseline signal generated by the sensor in the absence of illumination, accounting for global electronic bias and thermal noise. Although this correction does not fully eliminate fixed-pattern noise, it helps to suppress global dark-level artifacts and improves the accuracy of subsequent color calibration and holographic reconstruction.

### S6.2 Evaluation of Alternative Nonlinear Mapping Models

To justify the proposed architecture, we benchmarked the deep MLP against three competitive baselines under identical training conditions: standard 3rd-order polynomial regression, root-polynomial regression, and a shallow MLP (configured with a single hidden layer of 64 units and LeakyReLU activation). Figure S9 presents the comparative results using the ΔE metric, demonstrating that the proposed deep MLP significantly outperforms these baselines. Notably, polynomial-based methods exhibit limited capability in modeling the response in low-luminance regions (corresponding to the lower-left region in Figs. S9(b) and (d)). Although the shallow MLP partially mitigates these local artifacts, its limited capacity leads to inferior overall fidelity, exhibiting a 30% higher average ΔE compared to the deep MLP. In contrast, the proposed deep MLP achieves the lowest overall ΔE, effectively capturing the complex non-linear color mapping.

Under narrowband laser illumination, the camera's spectral response exhibits intrinsic complex local non-linearities, as evidenced by the sharp local structures in the topographic maps of Fig. S10. While baseline methods like polynomials or shallow networks are effective at modeling general nonlinear relationships, they typically lack the capacity to precisely characterize these specific sharp transitions without compromising global accuracy. Consequently, the proposed deep MLP provides the enhanced representational capacity necessary to resolve these fine-grained spectral variations more accurately than the baselines.

### S6.3 Camera Color Correction Using a Color Correction Matrix

To establish a baseline for camera color correction, we employ a linear CCM based on a standard color checker. This method is widely used to mitigate sensor-induced color distortions [Finlayson et al. 2015] and is integrated into our comparative pipeline, referred to as CITL+CCM.

In the CCM optimization process, the color checker (Datacolor SpyderCHECKR), consisting of 48 color patches, is first captured under uniform broadband illumination for stable calibration. Raw RGB values are extracted from the captured image and mapped to reference linear RGB values using a initialized 3×3 (CCM). The CCM is then optimized via gradient descent to minimize the color error between the corrected colors and the ground truth. The optimization process is illustrated in Fig. S11 and detailed in Algorithm S4.

This optimized CCM is subsequently applied to all captured holographic images during the CITL+CCM optimization process to correct camera-induced color distortions.

However, the CCM calibrated under broadband lighting conditions may perform poorly when applied to holographic images illuminated by narrow-band RGB lasers. As illustrated in Fig. S12, the same intended target color can result in different sensor RGB responses under broadband versus laser illumination due to the camera's characteristics and the narrowband nature of laser sources. This spectral mismatch introduces residual color distortions even after applying the CCM, limiting the color accuracy achievable under laser-based holographic displays. These findings highlight the limitations of conventional calibration methods and motivate our MLP-based perceptual color restoration approach, which explicitly accounts for spectral shifts and perceptual color discrepancies under laser-based narrowband illumination.

## S7 Additional Results

### S7.1 Additional Experimentally Captured Results

Figures S13–S15 present the complete set of experimentally captured 2D holographic reconstructions across all evaluated methods. These results span a diverse collection of 31 natural scenes, providing a comprehensive visual comparison of color and perceptual quality across methods. Figure S16 shows the captured 3D holographic image of a complex color scene.

### S7.2 Additional Simulation Results

Figures S17–S18 show the ablation study simulated holographic images under different ablation configurations across 31 scenes.

XXX:10 • Chen, C. et al.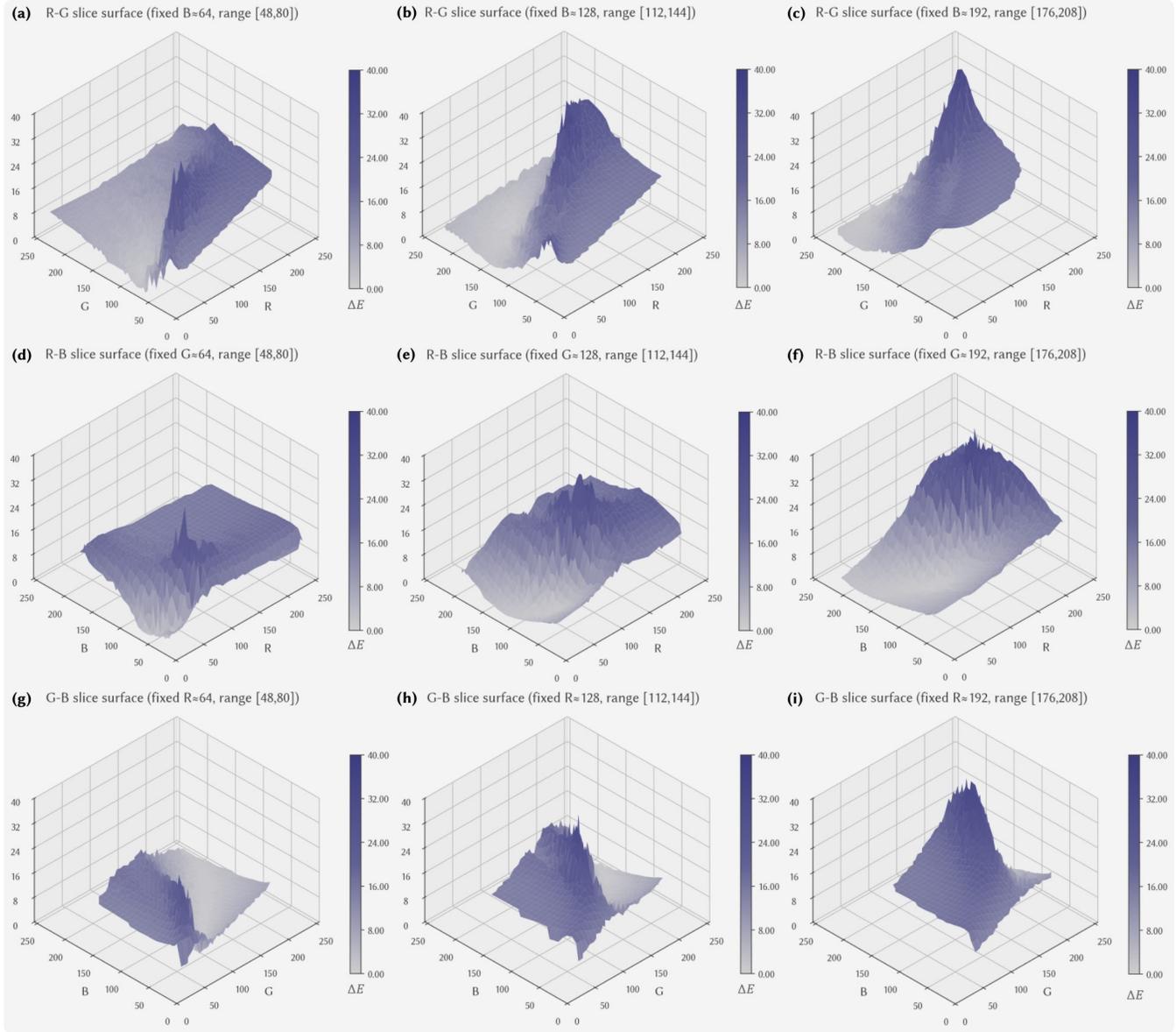

Fig. S10. Visualization of the camera's intrinsic response nonlinearity. This figure presents cross-sectional slices of the uncorrected color error ($\Delta E$) distribution (corresponding to the volumetric data in Fig. 5(a) of the main manuscript). The topographic maps visualize the error landscape projected onto different color planes: (a)-(c) the R-G plane, (d)-(f) the R-B plane, and (g)-(i) the G-B plane. The presence of irregular, sharp local structures (high-frequency variations) reveals the complex nonlinearity of the camera's raw response under narrowband illumination. These sharp transitions demonstrate why global low-order models (such as polynomials) are insufficient and justify the need for a deep network architecture with high representational capacity.

### S7.3 Additional User Study Results

Figure S19 reports the calculated Just-Objectionable-Difference (JOD) scores for all 31 scenes evaluated in the user study, offering a comprehensive assessment of perceptual color differences across various content. The color bar at the bottom indicates the grouping of scenes by color category.

ACM Trans. Graph., Vol. XX, No. X, Article XXX. Publication date: January 2026.

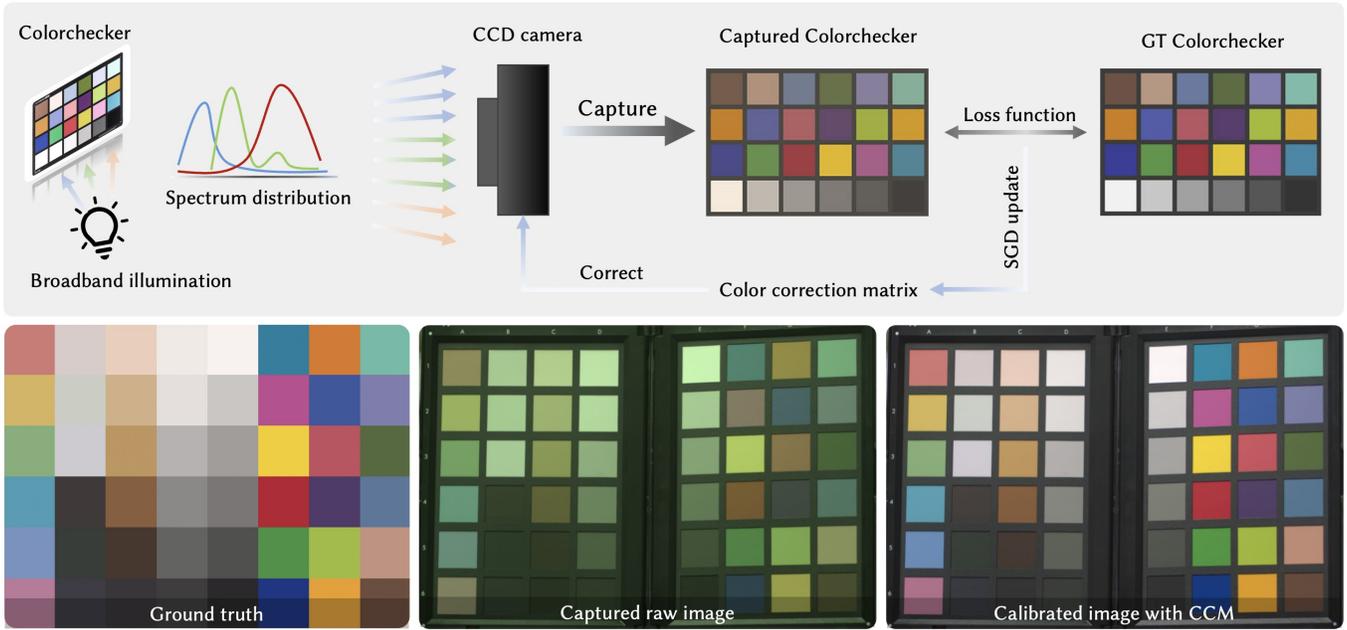

Fig. S11. Camera color correction using a CCM optimized from a standard color chart. The first row illustrates the CCM optimization process based on captured and ground truth colors. The second row compares camera-captured images without correction (left) and with the optimized CCM applied (right).

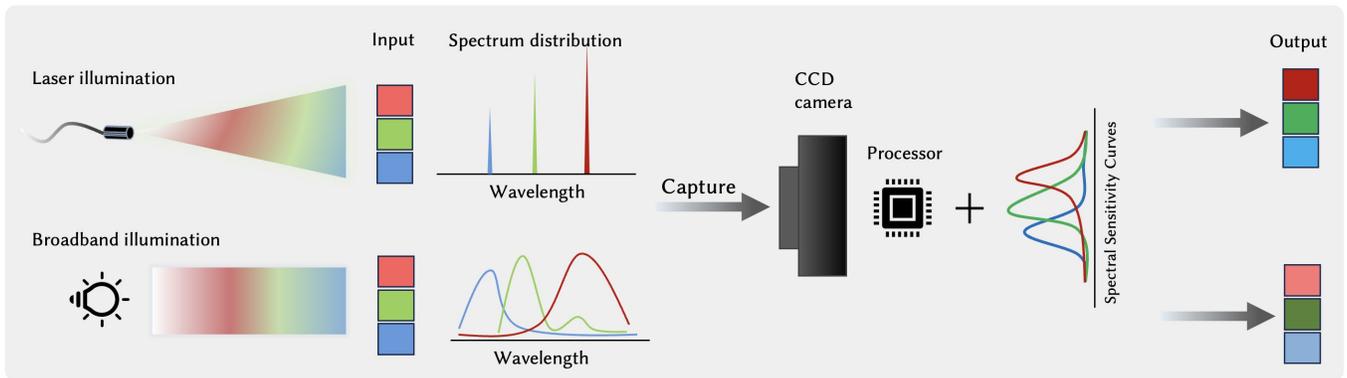

Fig. S12. Discrepancy in camera RGB responses under broadband vs. narrowband laser illumination. The same target color results in divergent sensor outputs, highlighting the need for color correction under the narrow-bandwidth illuminations.

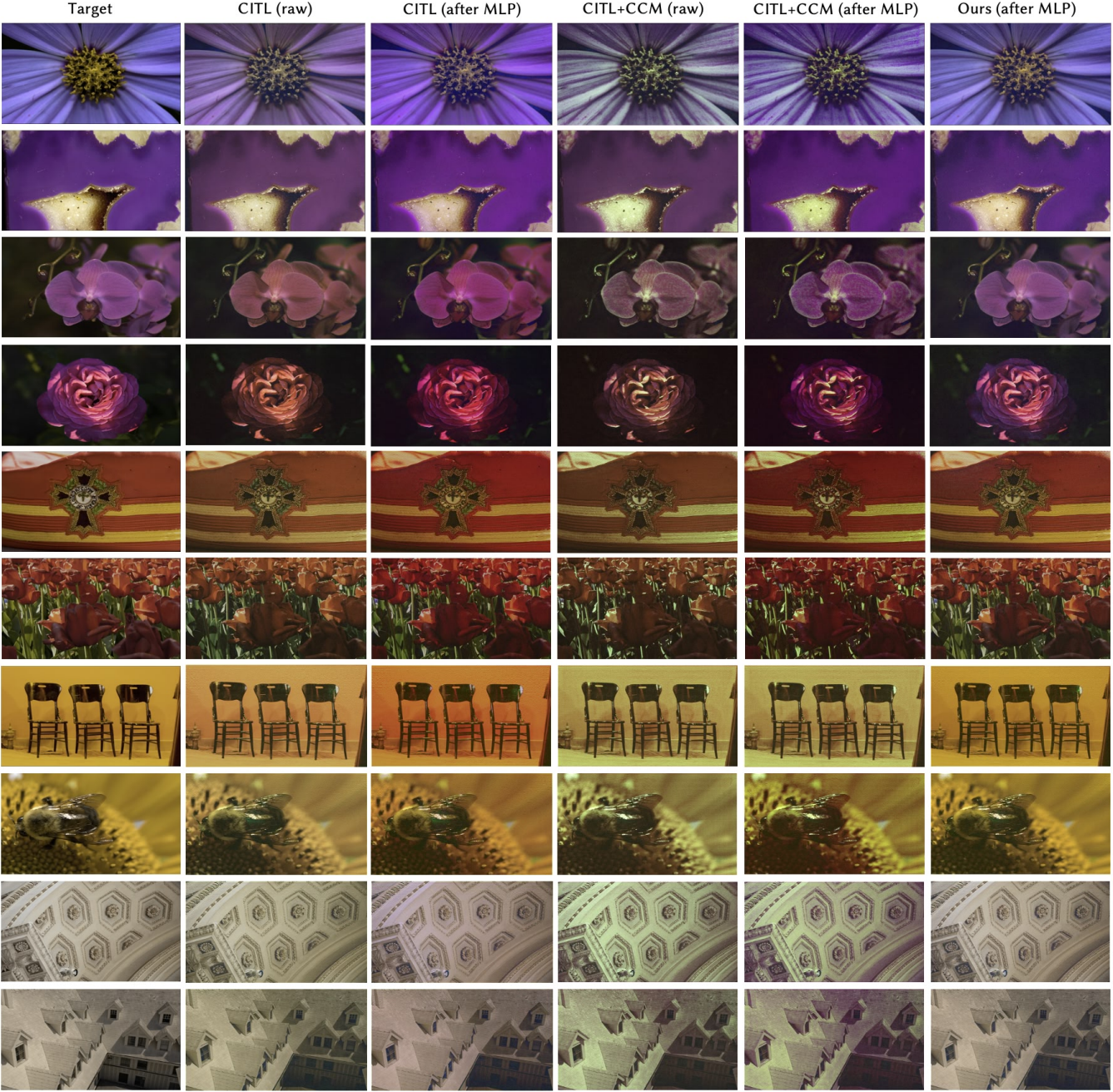

Fig. S13. Additional experimentally captured 2D holographic reconstructions from natural scenes (1/3).

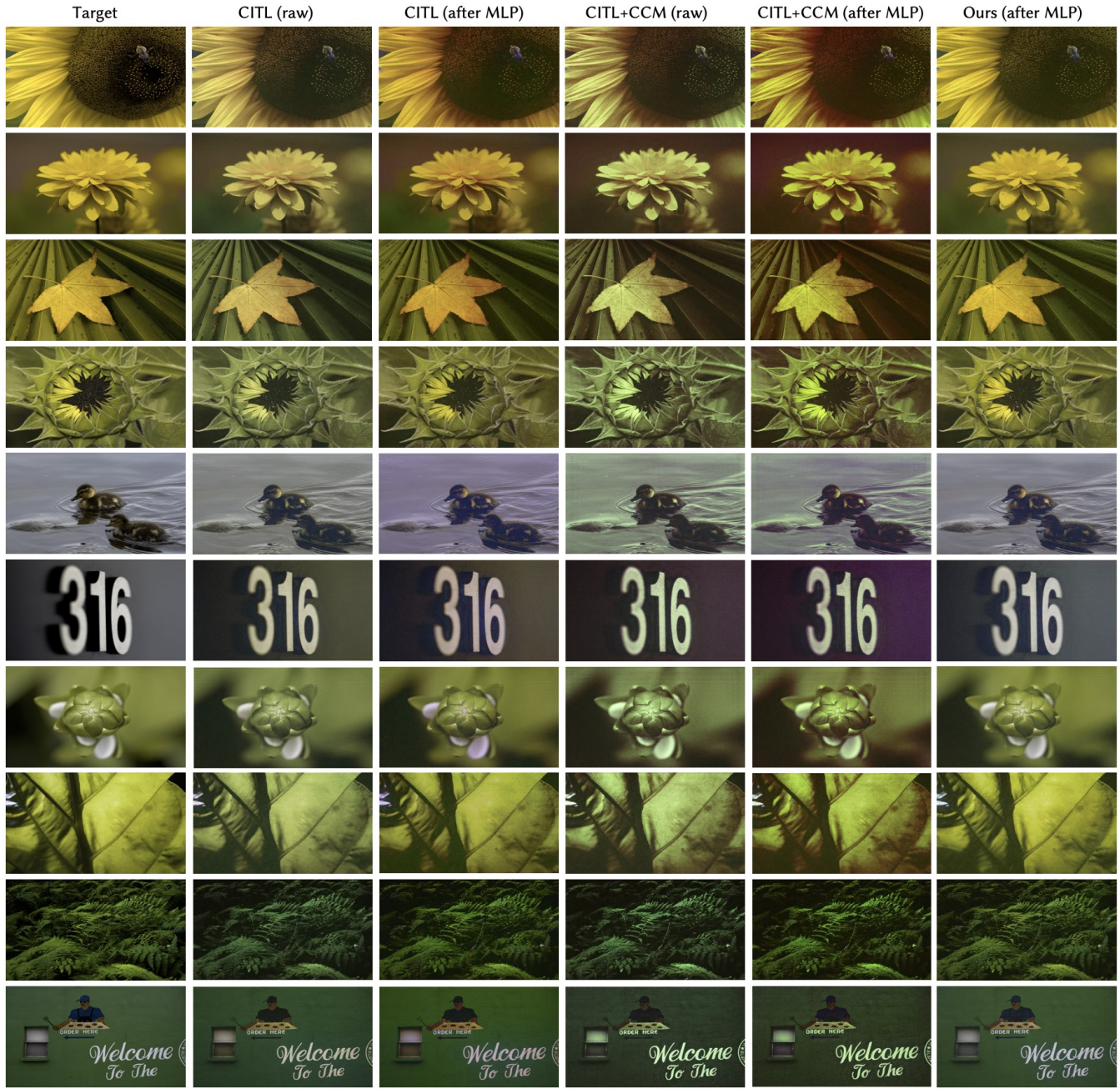

Fig. S14. Additional experimentally captured 2D holographic reconstructions from natural scenes (2/3).

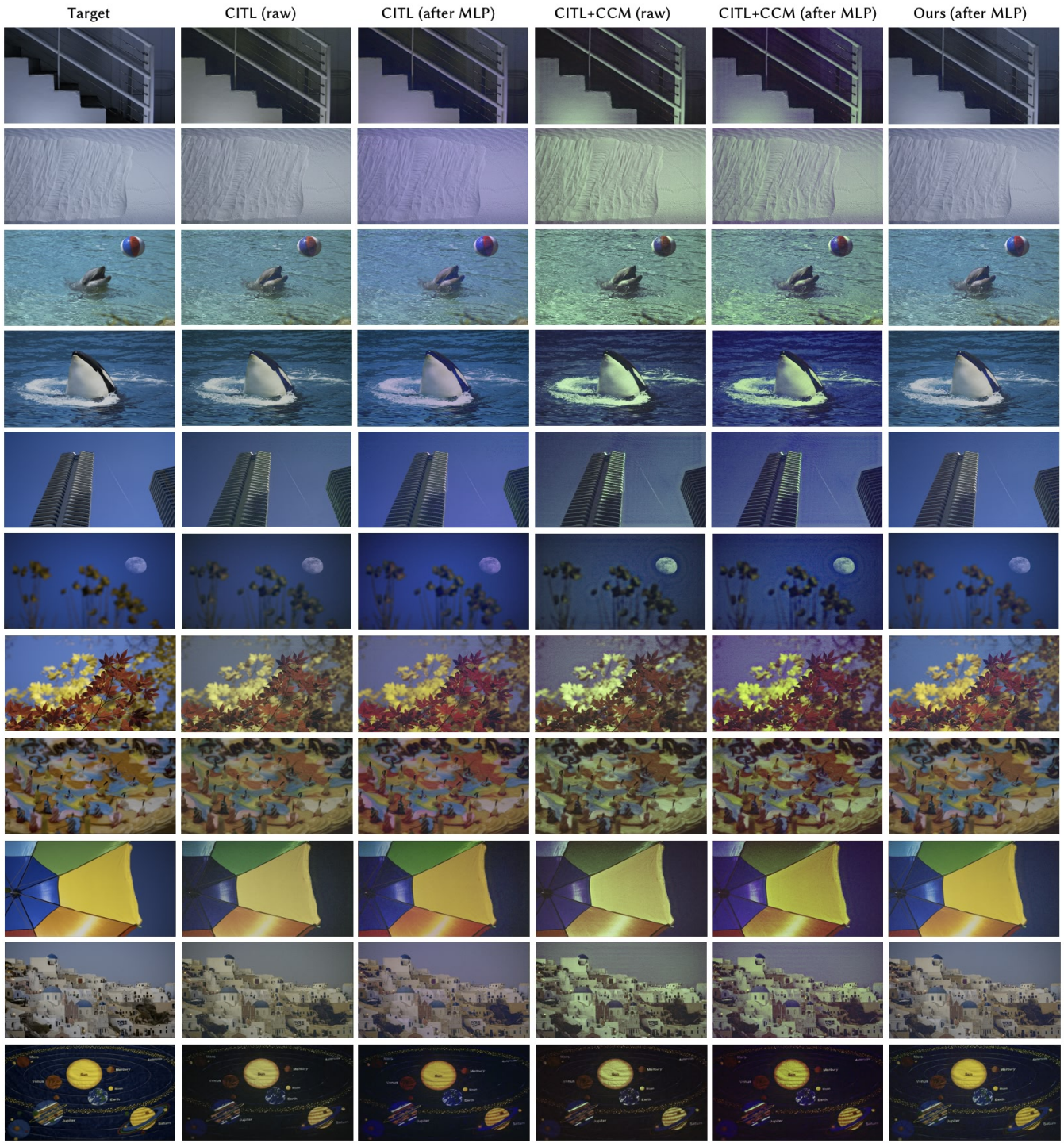

Fig. S15. Additional experimentally captured 2D holographic reconstructions from natural scenes (3/3).

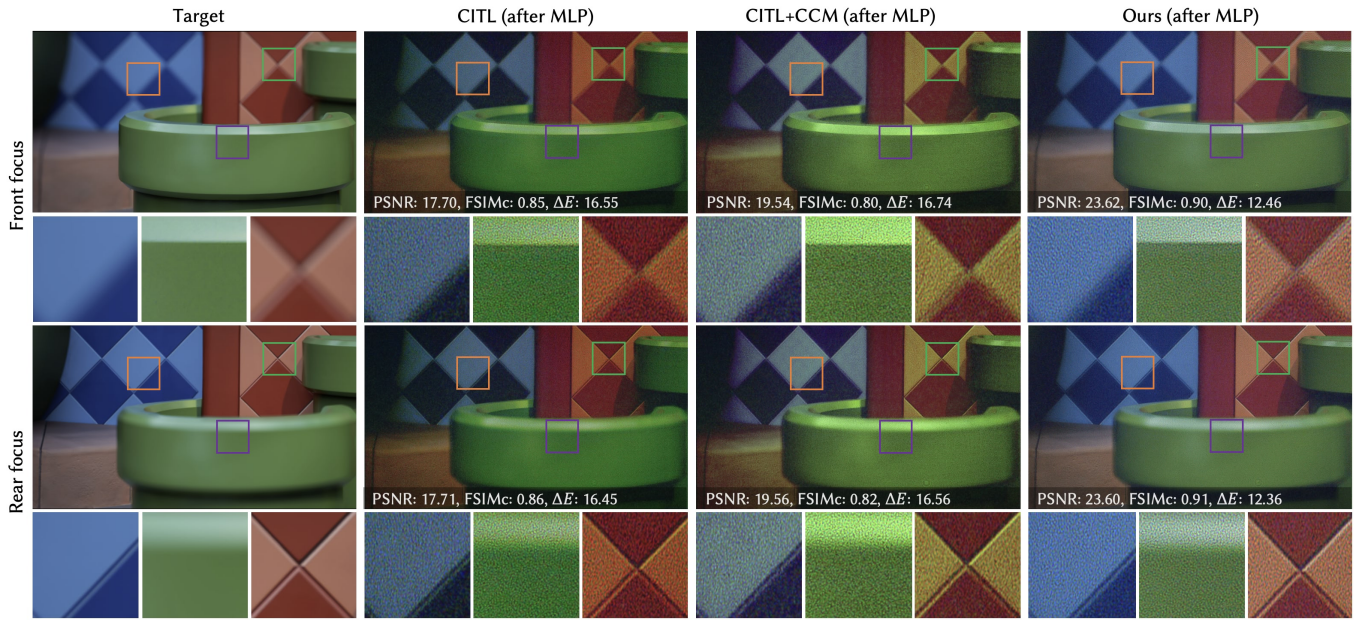

Fig. S16. Additional experimentally captured 3D holographic reconstructions for color scenes.

XXX:16 • Chen, C. et al.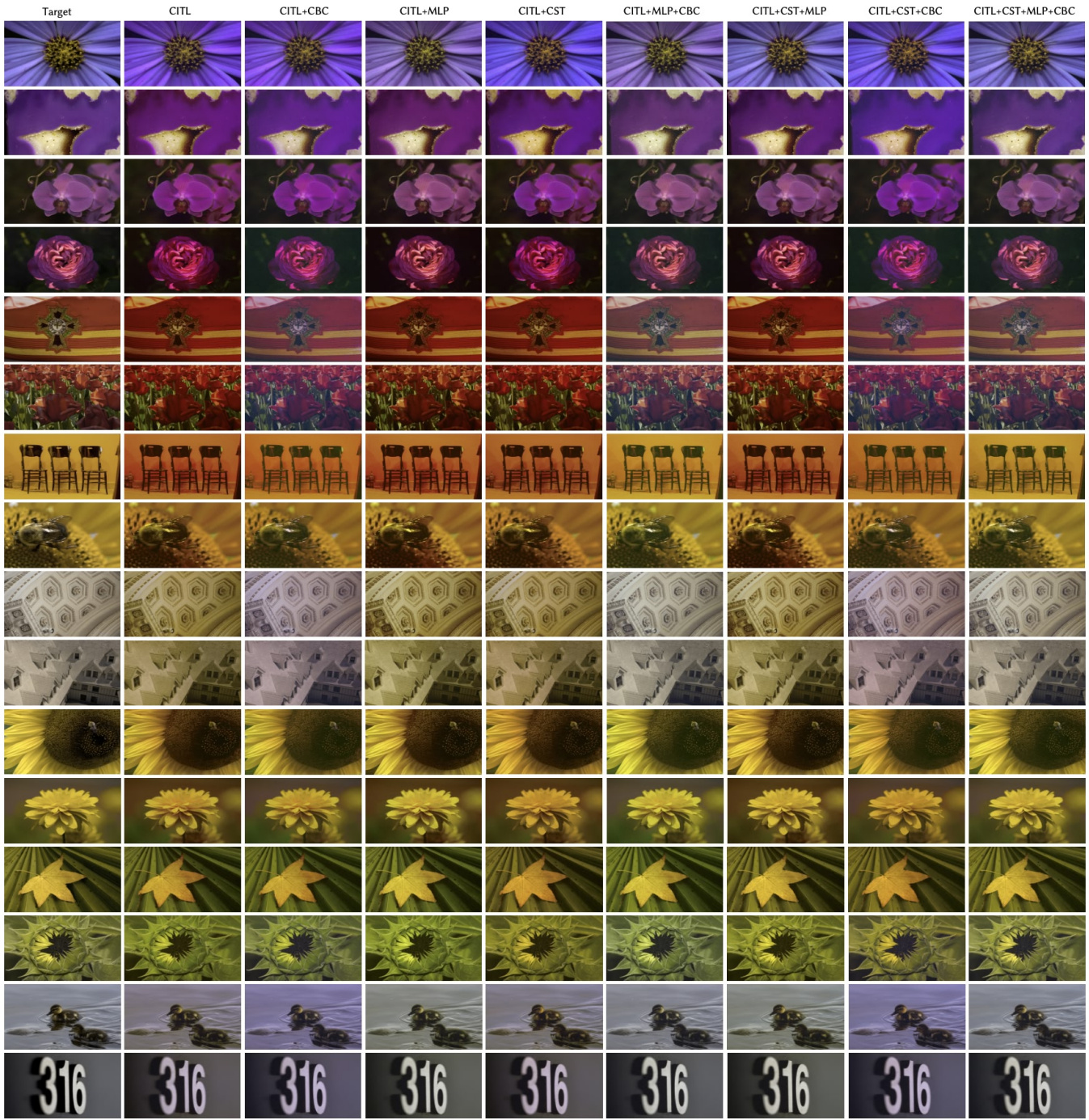

Fig. S17. Simulated holographic reconstructions for ablation study (1/2), illustrating the impact of different module combinations across diverse natural scenes.

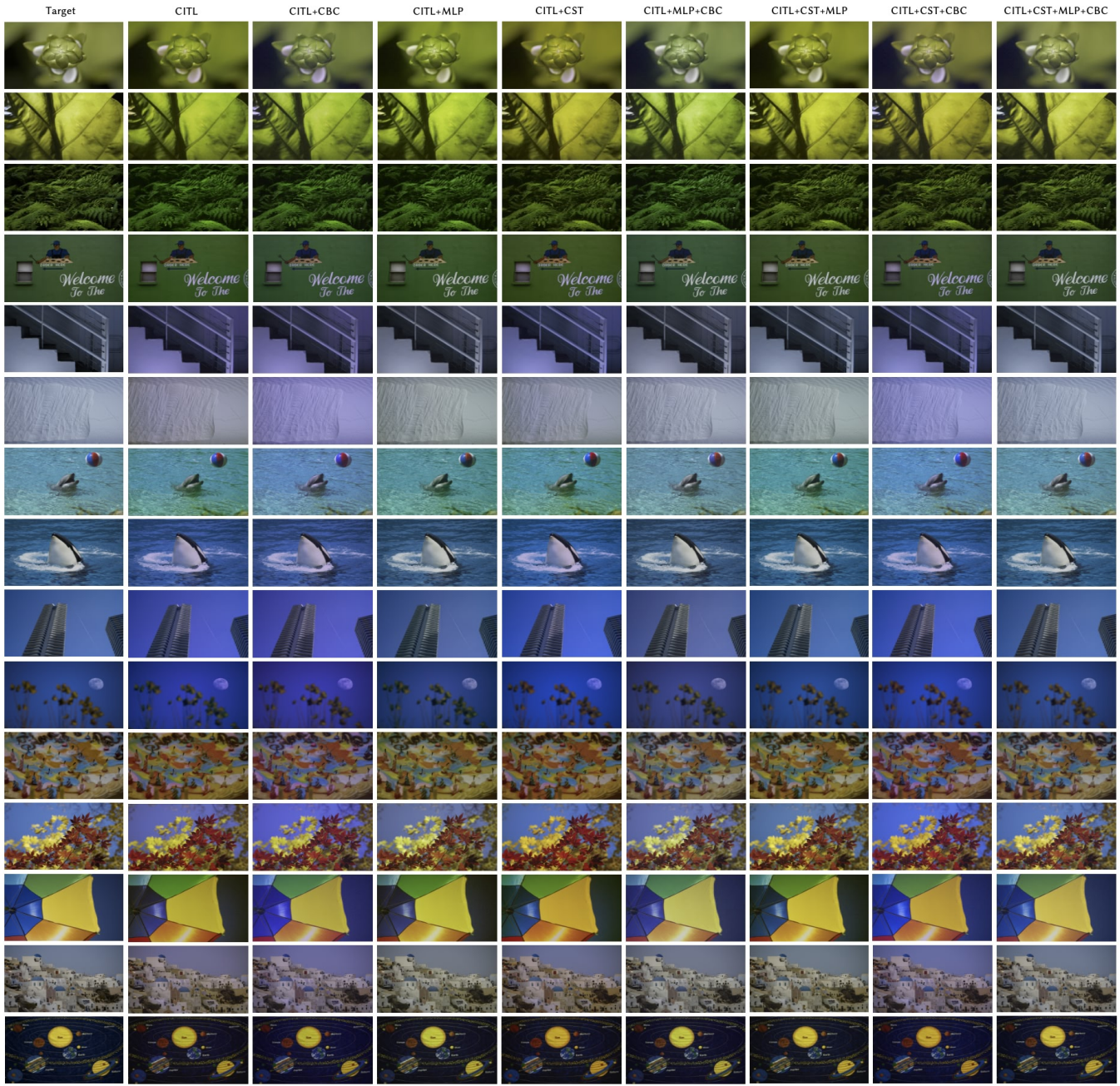

Fig. S18. Simulated holographic reconstructions for ablation study (2/2), further demonstrating the effects of individual components on color fidelity.

XXX:18 • Chen, C. et al.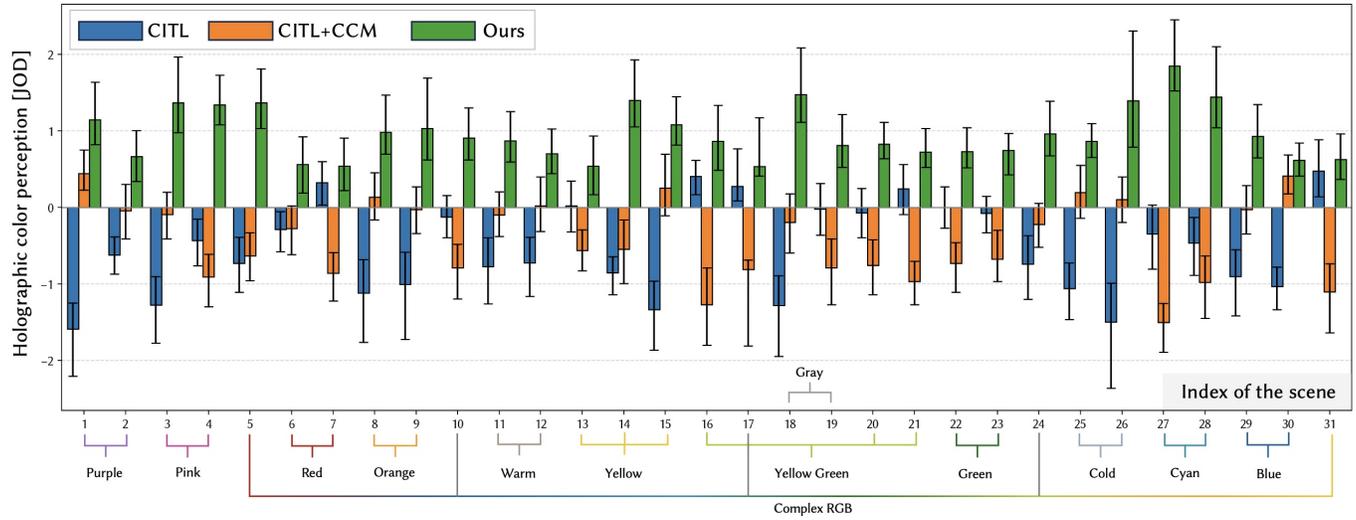

Fig. S19. JOD scores across all 31 scenes, reflecting perceptual color differences across color groups.

ACM Trans. Graph., Vol. XX, No. X, Article XXX. Publication date: January 2026.